\documentclass[aps,prd,eqsecnum,amsmath,amssymb,longbibliography,10pt,notitlepage]{revtex4-1}
\usepackage[utf8]{inputenc}
\usepackage[T1]{fontenc} 
\usepackage[colorlinks=true, linkcolor=blue, citecolor=teal]{hyperref}
\usepackage{bm}
\usepackage{latexsym,amssymb,amsmath,amsfonts}
\usepackage{graphicx}
\usepackage{physics}
\usepackage{float}
\usepackage{placeins}
\usepackage{footnotebackref}
\usepackage{multirow}
\usepackage[usenames]{xcolor}
\usepackage{soul}
\usepackage{makecell}
\usepackage{graphicx}
\usepackage{amssymb}  
\usepackage{amsmath}
\usepackage{rotating}
\usepackage{color}
\usepackage{ifthen}
\usepackage{slashed}
\usepackage{xspace}
\usepackage{tabularx}
\usepackage{float}
\usepackage[normalem]{ulem}
\def\be{\begin{equation}}
\def\ee{\end{equation}}
\def\ba{\begin{aligned}}
\def\ea{\end{aligned}}
\def\bi{\begin{itemize}}
\def\ei{\end{itemize}}
\def\ben{\begin{enumerate}}
\def\een{\end{enumerate}}
\def\i{\item{}}

\def\d{{\rm d}}
\def\llangle{\langle\!\langle}
\def\rrangle{\rangle\!\rangle}

\newcommand{\mb}[1]{\boldsymbol{#1}}

\def\cross{\times}

\newcommand{\Fmat}{\bm{\mathcal{F}}}
\newcommand{\lmax}{\ell_{\rm max}}
\newcommand{\lgwb}{\ell_{\rm max}^{\rm gwb}}
\newcommand{\lrec}{\ell_{\rm max}^{\rm rec}}
\newcommand{\lres}{\ell_{\rm max}^{\rm res}}
\newcommand{\lnn}{\ell_{\rm max}^{\rm nn}}
\newcommand{\lNpair}{\ell_{\rm max}^{N_{\rm pair}}}
\newcommand{\leff}{\ell_{\rm eff}}
\newcommand{\barl}{{\bar{\ell}}}
\newcommand{\barm}{{\bar{m}}}

\newcommand{\hatom}{\hat{\Omega}}

\begin{document}

\title{Addressing leakage and mode suppression in angular power spectrum estimation \\
for gravitational-wave backgrounds using pulsar timing arrays}

\author{Deepali Agarwal}
\email{deepali.agarwal@utrgv.edu}
\affiliation{Department of Physics and Astronomy, University of Texas Rio Grande Valley, One West University Boulevard, Brownsville, Texas 78520, USA}

\author{Joseph D. Romano}
\email{joseph.romano@utrgv.edu}
\affiliation{Department of Physics and Astronomy, University of Texas Rio Grande Valley, One West University Boulevard, Brownsville, TX 78520, USA}

\author{Yacine Ali-Ha\"imoud}
\email{yah2@nyu.edu}
\affiliation{Center for Cosmology and Particle Physics, Department of Physics, New York University, New York, New York 10003, USA}

\author{Tristan L. Smith}
\email{tsmith2@swarthmore.edu}
\affiliation{Department of Physics and Astronomy, Swarthmore College, 500 College Avenue, Swarthmore, Pennsylvania 19081, USA}

\begin{abstract}
Mapping gravitational-wave background (GWB) anisotropy with pulsar timing arrays (PTAs) is affected by harmonic space mode suppression and mode coupling arising from an array's nonuniform sky response. 
Due to computational limitations, one must truncate spherical harmonic expansions at a finite multipole for recovery, $\lrec$, often chosen to equal $\lNpair\equiv {\rm int}\left[\sqrt{N_{\rm pair}}-1\right]$, where $N_{\rm pair}\equiv N_{\rm psr}(N_{\rm psr}-1)/2$ is the number of distinct pulsar pairs in an array comprised of $N_{\rm psr}$ pulsars.
This choice is motivated by the counting argument that cross-correlations provide at most $N_{\rm pair}$ independent pieces of information for a single frequency bin. 
Here, we explicitly obtain the multipole value (denoted $\lres$) that approximates the maximum informative angular scale of an array of pulsars. This value is defined by the requirement that a spherical harmonic expansion out to $\lres$ approximately span the space of ``observable skies'' extracted by the array, which is encoded in the $N_{\rm pair}$ eigenmaps of the Fisher information matrix. 
The value of $\lres$ depends on specifics of the PTA configuration.
We also explicitly show that GWB power contained in multipoles $\ell\gtrsim \lres$ do not significantly affect analyses that use expansions out to $\lres$, due to the mode suppression (i.e., low-pass filtering) induced by the PTA response to the GWB.
However, truncating spherical harmonic expansions at $\lrec< \lres$ leads to leakage of small-scale angular power from multipoles $\lrec<\ell\le\lres$. 
Nonetheless, even if we use $\lres$ for recovery, the standard frequentist estimator of the angular power spectrum components $C_{\ell}$ is still biased due to modes that are not observable by the array. 
Although we can partially debias the standard estimator---improving its agreement with an injected angular power spectrum---this reduction in bias comes at the expense of an increase in variance, with especially large variances arising for poorly constrained modes, $\ell \gg\leff$.
In summary, the results of our analyses strongly advocate for: (i) using $\lres$ for PTA analyses involving spherical harmonic expansions, and (ii) using the debiased standard estimator for $C_\ell$ recovery, but only out to multipoles $\ell<\leff(\ll\lres)$ corresponding to sufficiently constrained modes.
\end{abstract}

\maketitle

\section{Introduction}
\label{s:intro}

Recent pulsar timing array (PTA) observations have opened a new window onto the Universe by providing weak to compelling evidence of low-frequency ($\sim$nanoHz) gravitational waves (GWs)~\cite{10.1093/mnras/stae2571,Reardon_2023,Xu_2023,Agazie_2023,EPTA_InPTA_23}. 
These observations are consistent with the predicted correlations among pulsar pairs expected from an isotropic gravitational-wave background (GWB). 
However, the origin of the signal remains uncertain due to uncertainties in source and noise modeling and the limited sensitivity of current observations~\cite{NANOGrav:2023hvm}. 
Although, no significant anisotropy has been detected so far~\cite{Taylor:2015udp,NANOGrav:2023tcn,Grunthal:2024sor,Chen:2026mid}, anisotropies in the GWB are expected since GW sources of astrophysical origin (e.g., a population of supermassive black-hole binaries in the centers of merging galaxies) trace the large-scale structure of the Universe~\cite{Mingarelli_2017,Sato-Polito:2023spo,Gardiner:2023zzr,Semenzato:2024mtn,Sah:2024oyg,Lin:2026owz,Lin:2026qvc}.

Assuming the cosmological principal, these anisotropies are expected to be statistically isotropic and Gaussian. 
When modeling an anisotropic GWB, it is often convenient to expand the GW power on the sky, $P(\hat\Omega)$, in terms of spherical harmonics (SpHs), $Y_{\ell m}(\hat\Omega)$, and characterize the anisotropy by its angular power spectrum, whose form depends on the underlying source population~\cite{Semenzato:2024mtn,Sah:2024oyg,Sah:2024etc}.
Therefore, an inaccurate estimation of the angular power spectrum may lead to biased astrophysical and cosmological inferences.

Any attempt to infer the angular power spectrum of a stochastic GWB using PTA data must contend with the fact that the response of the pulsar array to the background provides only partial information of the distribution of GW power on the sky.
More specifically, for a PTA comprised of $N_{\rm psr}$ pulsars, the data contain at most $N_{\rm pair}\equiv N_{\rm psr}(N_{\rm psr}-1)/2$ independent pieces of information per frequency bin, corresponding to the $N_{\rm pair}$ distinct cross-correlations of the timing residuals between the pulsar pairs. 
One might therefore choose $\lNpair$ to be the largest multipole, for which the total number of SpH modes $\{\ell, m\}$ satisfies $(\lNpair+1)^2\le N_{\rm pair}$ (so that $\lNpair\sim N_{\rm psr}$)~\cite{Semenzato:2025sqc,Domcke:2025esw}. 
(See Table~\ref{t:lmax_defs} for definitions of SpH multipoles corresponding to different angular scales.)

However, as demonstrated in~\cite{Semenzato:2025sqc}, this choice leads to excessive leakage of small-scale angular power into the reconstructed angular power spectrum. 
Also, as hinted in~\cite{Semenzato:2025sqc}, this excess leakage is progressively suppressed as $\lrec$ is increased.
Reference~\cite{Grunthal2026} has shown that the above counting-based truncation also affects point-source reconstruction using a SpH expansion. The authors determine the value of $\lmax$ that yields the smallest reconstructed point-source size. 
Another counting-based limit, originally derived for phase-coherent mapping~\cite{Romano:2016dpx}, has also been adopted for intensity mapping and applied to PTA data, $\lrec\sim \sqrt{N_{\rm psr}}$~\cite{Boyle:2010rt,Romano:2016dpx,Pol:2022sjn,NANOGrav:2023tcn,Grunthal:2024sor}. 
This choice truncates the reconstruction at an even smaller multipole, thus potentially enhancing the leakage-induced bias.

In this article, we use the Fisher information matrix to identify the multipoles for which the PTA response retains significant information. 
The Fisher information matrix depends explicitly on the PTA response and the pulsar noise properties, thereby naturally accounting for the geometry of the array~\cite{Ali-Haimoud:2020ozu,alihaimoud2021,Pol:2022sjn,Grunthal:2024sor,Grunthal2026}. 
Under the assumption of a Gaussian likelihood for the pulsar cross-correlations, the Fisher information matrix naturally appears both as the convolution matrix relating the true sky to the observed map, and also as the inverse of the covariance of the anisotropy estimators~\cite{Mitra:2007mc,Thrane:2009fp,alihaimoud2021,Pol:2022sjn}. 
The angular information accessible to a PTA is encoded in the $N_{\rm pair}$ eigenmaps of the Fisher information matrix, which span the observable subspace of the GW sky~\cite{Ali-Haimoud:2020ozu,alihaimoud2021}. 

As we show in Sec.~\ref{s:lres}, we typically are not be able to reconstruct with sufficient accuracy a sufficient number of the eigenvectors of the Fisher information matrix if we choose $\lNpair$ for cutoff. 
However, if the reconstruction cutoff $\lrec$ is chosen to be sufficiently large, the SpHs up to $\lrec$ provide an approximately complete basis for the observable subspace for the PTA. 
We define the corresponding threshold multipole, $\lres$, as the minimum value of $\lmax$ for which a sufficient fraction (e.g., 90\%) of the $N_{\rm pair}$ eigenvectors of the Fisher information matrix can be reconstructed to sufficient accuracy (e.g., 95\% match) using SpHs up to $\lres$. 
This value of $\lmax$ allows us to approximately capture the smallest angular scales accessible to the array. 
Consequently, the excess leakage can be removed by using $\ell_{\rm max}^{\rm rec}\ge \ell_{\rm max}^{\rm res}$. 
We verify these conclusions using numerical simulations and analytical calculations detailed in Secs.~\ref{s:demos} and \ref{s:other-PTA-configs}. 
We also show that the threshold value $\lres$ depends on specifics of the PTA configuration.

In Secs.~\ref{s:lres} and ~\ref{s:demos}, we also explain how leakage is due to mode coupling, which manifests itself as a nonspherically symmetric Fisher matrix in the pixel basis.
In addition, we point out how mode suppression limits the leakage of small-scale angular power from the true GWB power distribution.
This means that even if the GWB has power out to $\lgwb$ of order several hundred or more (assuming a constant angular power spectrum), the response of a PTA acts much like a low-pass filter of the GWB power, effectively restricting the observable multipoles to $\ell\le\lres$.
Finally, we show that even if we use a sufficiently large $\lrec$, the standard frequentist estimator of the angular power spectrum, $C_\ell$, is still biased due to modes that are not observable by the array. 
It turns out that we can debias the standard estimator, reducing its bias at the expense of an increase in variance.
But the increase in variance becomes especially large for the most poorly constrained modes.

The rest of the paper is organized as follows:
In Sec.~\ref{s:preliminaries}, we introduce the mathematical formalism needed to discuss the frequentist cross-correlation estimators of the angular power spectrum $C_\ell$ used for our analyses.
Sections \ref{s:lres} and \ref{s:demos} presents the main results, which we have already highlighted above.
In Sec.~\ref{s:other-PTA-configs}, we show how these results depend on specifics of the PTA configuration.
Finally in Sec.~\ref{s:summary}, we conclude with a brief summary and discussion of possible extensions of our work.
Appendix~\ref{s:stats_prop} contains some mathematical details regarding the statistical properties of the various estimators used in this paper.
Appendix~\ref{app:noBiasNoCoupling} demonstrates that, in the absence of mode coupling, unmodeled GWB angular scales do not bias the angular power spectrum estimator.
And Appendix~\ref{s:condition_number} discusses the impact of the choice of the condition number for regularizing matrix inversion on $C_\ell$ recovery.

\begin{table*}[tb!]
    \centering
    \renewcommand{\arraystretch}{1.5} 
    \begin{tabular}{c|l}
    \hline
    Notation & $\ $ \hspace{2in}Definition 
    \\
    \hline
    $\lgwb$ & maximum multipole for the true distribution of GW power on the sky (unknown)
    \\
    $\lrec$ & maximum multipole used for the recovered distribution of GW power (our signal model)
    \\
    $\lres$ & multipole corresponding to the maximum angular resolution of a PTA (configuration and accuracy dependent)
    \\
    $\leff$ & multipole corresponding to the effective angular resolution of a PTA (configuration and accuracy dependent)
    \\
    $\lNpair$ & multipole for which the total number of SpH modes $\{\ell,m\}$ is $\le N_{\rm pair}$, i.e., $\lNpair={\rm int}[\sqrt{N_{\rm pair}}-1]$
    \\
    $\lnn$ & multipole defined by the angular separation $\delta_{\rm nn}$ (in degrees) of nearest-neighbor pulsar pair, i.e., $\lnn=180^\circ/\delta_{\rm nn}$
    \\
    \hline
    \hline
    \end{tabular}
    \caption{Multipoles corresponding to different angular scales discussed in this paper.}
    \label{t:lmax_defs}
\end{table*}

%%%%%%%%%%%%%%%%%%%%%%%%%%%%
\section{Mathematical preliminaries}
\label{s:preliminaries}

\subsection{Map-making using PTA data}

As mentioned in the introduction, we are interested in searching for an anisotropic GWB.
Assuming that an unpolarized GWB signal can be characterized as a stationary Gaussian process, the Fourier coefficients of the GW strain satisfy~\cite{Romano:2016dpx}:
\be 
\langle \tilde{h}_A(f,\hatom)\,\tilde{h}_{A'}^*(f',\hatom') \rangle  = \,\delta_{AA'}\,\delta^{2}(\hatom,\hatom')\,\delta(f-f')P(f,\hatom)\,,
\label{eq:strain_expectation}
\ee
where $\langle\ \rangle$ denotes ensemble average over different GW strain (and, if present, noise) realizations, and $P(f,\hatom)$ is a one-sided power spectral density (PSD), which is the primary quantity probed by GW experiments. 
For a broadband source, the frequency and direction dependence of the PSD is often factorized as 
\be
P(f,\hatom)=H(f)\,P(\hatom)\,.
\ee
Since our focus is exclusively on anisotropy, we set $H(f)=1$ in the following discussion, which is equivalent to estimating only $P(\hat\Omega)$.

For an extended distribution of GWB power on the sky, it is convenient to decompose the angular dependence of $P(\hatom)$ in a SpH basis as
\be 
\label{eq:sph_decompose}
P(\hatom) = \sum_{\barl=0}^{\ell_{\rm max}^{\rm gwb}}\sum_{\barm=-\barl}^{\barl}\,P_{\barl\barm } \,Y_{\barl \barm}(\hatom)\,,
\ee
where $P_{\barl\barm}$ are the SpH coefficients of $P(\hat\Omega)$ and $Y_{\barl\barm}(\hatom)$ are the corresponding basis functions. 
We have assumed that the GW power has support on angular scales $0\leq\barl\leq\ell_{\rm max}^{\rm gwb}$ (in principle, $\lgwb$ can be arbitrarily large). 
For a statistically isotropic GWB (which has no preferred direction), and assuming that the monopole has been perfectly subtracted, the SpH coefficients are fully characterized by the angular power spectrum components $C_\ell$ defined via
\be \label{eq:stats_Iso}
\langle P_{\barl \barm} \rangle_P =0 \,,\qquad
\langle P_{\barl \barm} P^*_{\barl' \barm'} \rangle_P = C_\barl\, \delta_{\barl \barl'}\,\delta_{\barm \barm'}\,,
\ee
where $\langle\ \rangle_P$ denotes ensemble average over different $P_{\barl \barm}$ realizations corresponding to a fixed set of $C_\barl$.

In PTA experiments, the SpH components and angular power spectrum of an anisotropic GWB are estimated from cross-correlations, $\rho_{ab}$, of timing residuals between distinct pulsar pairs, labeled by $a<b$. 
The expectation value of the measured cross-correlation $\rho_{ab}$ is related to $P(\hatom)$ via~\cite{Pol:2022sjn,NANOGrav:2023tcn}:
\be \label{eq:rho_ab_expectation}
    \langle \rho_{ab} \rangle 
    =\int {\rm d}^2\hatom\>
    \gamma_{ab}(\hatom)\,P(\hatom)
    =\sum_{\barl=0}^{\ell_{\rm max}^{\rm gwb}}\sum_{\barm=-\barl}^{\barl}\gamma_{ab,\barl \barm}\,P_{\barl \barm}\,,
\ee
where
\be
\gamma_{ab}(\hatom) \equiv \sum_{A=+,\cross}F^A(\hat{p}_a,\hatom)\,F^A(\hat{p}_b,\hatom)\,,
\qquad
\gamma_{ab, \barl \barm} \equiv \,\int \d^2\hatom\>\gamma_{ab}(\hatom)\,Y_{\barl \barm}(\hatom)\,.
\label{e:ORFmap}
\ee
Here $F^A(\hat{p}_a,\hatom)$ denotes the response of pulsar $a$ to a GW with polarization $A$ propagating in direction $\hatom$. The functions $\gamma_{ab}(\hatom)$ can be written explicitly as simple functions of the dot products between $\hat{p}_a$, $\hat{p}_b$, and $\hatom$, as shown in~\cite{Ali-Haimoud:2020ozu}. The quantity $\gamma_{ab,00}$ corresponds to the Hellings-Downs correlation~\cite{Hellings-Downs:1983}, which depends only on the separation angle between $\hat{p}_a$ and $\hat{p}_b$.
(Here, $\hat p_a$ denotes the unit vector pointing from Earth to pulsar $a$.) It what follows, we will use the nomenclature of~\cite{Ali-Haimoud:2020ozu, alihaimoud2021} and refer to the $\gamma_{ab}(\hatom)$ as pairwise timing response (PTR) maps. 

To reconstruct the SpH coefficients $P_{\ell m}$ from the data, we assume that the signal has support for multipoles $0\leq\ell\leq \ell_{\rm max}^{\rm rec}$. (Note that we will use unbarred indices $(\ell,m)$ to denote SpH coefficients for the recovered signal, as opposed to barred indices $(\barl,\barm)$ when referring to the true GWB.)
Assuming that the cross-correlations are described by a Gaussian distribution, the likelihood function is maximized with respect to $P_{\ell m}$ when \cite{Mitra:2007mc,Thrane:2009fp,Pol:2022sjn,NANOGrav:2023tcn,Grunthal:2024sor,Grunthal2026}:
\be \label{eq:clean_estimator}
    \hat{P}_{\ell m} =\sum_{\ell' m'} (\mathcal{F}^{-1})_{\ell m, \ell' m'}\, X_{\ell' m'}\,,
\ee
where the ``dirty map'' $X_{\ell m}$ and SpH coefficients $\mathcal{F}_{\ell m, \ell' m'}$ of the Fisher information matrix, $\Fmat$, are defined via
\be \label{eq:dirty_fisher}
X_{\ell m}\equiv \sum_{a<b} \frac{\gamma^*_{ab,\ell m}\,\rho_{ab}}{\sigma_{ab}^2}\,,\qquad
\mathcal{F}_{\ell m, \ell' m'} \equiv \sum_{a<b} \frac{\gamma^*_{ab,\ell m}\,\gamma_{ab,\ell' m'}}{\sigma_{ab}^2}\,.
\ee
Here, $\sigma^2_{ab}$ denotes the variance of the cross-correlations $\rho_{ab}$. 
The noise covariance of the ``clean map'' estimator $\hat P_{\ell m}$ is given by $\mathcal{F}^{-1}_{\ell m, \ell' m'}$.

We note that \eqref{eq:dirty_fisher} implicitly assumes the \emph{weak-signal limit}, for which the timing noise is dominated by intrinsic pulsar noise.
This implies, in particular, that the covariance matrix of the cross-correlations $\rho_{ab}$ is diagonal. In general, a non-negligible GWB (in particular its monopole, which should dominate over anisotropies) would imply nonzero off-diagonal correlations between different pulsar pairs, involving the Hellings and Downs function, see e.g.,~\cite{Ali-Haimoud:2020ozu}. If we are able to measure the monopole sufficiently accurately, then this monopole-dependent, nondiagonal covariance matrix would replace $\sigma^2_{ab}$ in the definition \eqref{eq:dirty_fisher} of the dirty map and Fisher matrix, see e.g.,~\cite{Grunthal:2024sor,Semenzato:2025sqc}.
We will not consider this particular modification of \eqref{eq:dirty_fisher} for our analyses.

Finally, we note that, in practice, the Fisher matrix $\Fmat$ is ill-conditioned whenever the number of modes we are trying to recover is greater than the $N_{\rm pair}$ cross-correlation measurements that we have to work with.
For this case, $\Fmat$ must be regularized 
to obtain a pseudoinverse $\Fmat^{+}$. 
The clean map estimator \eqref{eq:clean_estimator} then becomes~\cite{Thrane:2009fp,Grunthal:2024sor,Semenzato:2025sqc,Grunthal2026} 
\be \label{eq:pseudo_clean_estimator}
    \hat{P}_{\ell m} =\sum_{\ell' m'} {\mathcal{F}}^{+}_{\ell m, \ell' m'}\, {X}_{\ell' m'}\,.
\ee
In this article, we employ a singular-value-decomposition (SVD)-based regularization scheme. 
The Fisher matrix is decomposed as 
\be\label{e:SVD_1}
\Fmat = \bm{U}\cdot\bm{\Sigma}\cdot \bm{V}^\dagger\,,
\ee
where $\bm{U}$ and $\bm{V}$ are unitary matrices, and $\bm{\Sigma}$ is a diagonal matrix with elements $\Sigma_{ij}=s_i\,\delta_{ij}$, with $s_i$ denoting the singular values. 
We note that for a square Hermitian matrix (as is the case for the Fisher matrix), the $\bm{U}$ and $\bm{V}$ matrices coincide and the SVD decomposition reduces to an eigenvalue-eigenvector decomposition.
To construct the pseudoinverse, we introduce a condition number threshold, $\kappa$. The pseudoinverse is defined as 
\be\label{e:SVD_2}
\Fmat^+ = \bm{V}\cdot\bm{\Sigma}^+\cdot \bm{U}^\dagger\,,
\ee
where 
\be\label{e:SVD_3}
(\Sigma^{+})_{ij}
=\delta_{ij}\left\{
\begin{array}{cl}
 1/s_i    & {\rm for}\ s_i > {\rm max}[s_i]/\kappa  \\
 0    & {\rm for}\ s_i \le {\rm max}[s_i]/\kappa
\end{array}
\right.\,.
\ee
In other words, $\bm{\Sigma}^+$ contains the inverse of the singular values for the retained modes, while modes with smaller singular values are set to zero for $\bm{\Sigma}^+$.

The necessity of using the pseudoinverse $\Fmat^{+}$ instead of true inverse $\Fmat^{-1}$ of the Fisher matrix implies that the $\hat P_{\ell m}$ estimators are biased~\cite{Thrane:2009fp,Grunthal:2024sor,Semenzato:2025sqc}, as shown explicitly in Appendix~\ref{s:stats_prop}, see \eqref{eq:clean_expectation} and \eqref{e:K}. 
This bias propagates to the standard frequentist estimator, $\hat C_\ell$, of the angular power spectrum $C_\ell$, as we will show in the following subsection. The bias and variance of these estimators depend on the condition number threshold: higher thresholds reduce the bias but increase the variance as we discuss in Appendix~\ref{s:condition_number}.

\subsection{Standard estimator for the angular power spectrum}

Given the clean map estimators $\hat P_{\ell m}$ defined in \eqref{eq:pseudo_clean_estimator}, we can then define an estimator of the angular power spectrum~\cite{Thrane:2009fp,Pol:2022sjn,NANOGrav:2023tcn,Semenzato:2025sqc}
\be \label{eq:Cl_estimator_1}
    \ba
    \hat{ C}_{\ell} &\equiv\frac{1}{2\ell+1}\, \sum_{m} |\hat{P}_{\ell m}|^2-N_{\ell}\,.
    \ea
    \ee
Note that we need to subtract a noise bias term~\cite{Thrane:2009fp} 
\be
\label{e:calN-main}
N_{\ell}
\equiv \frac{1}{2\ell + 1}
\sum_{\bar m}
{\mathcal{N}}_{\ell m,\ell m}\,,
\qquad
{\mathcal{N}}_{\ell m,\ell'm'}\equiv
(\Fmat^+\cdot\Fmat\cdot\Fmat^{+\dagger})_{\ell m,\ell' m'}\,,
\ee
which is derived in Appendix~\ref{s:stats_prop}.
In an ideal experiment, one would like to reconstruct as many modes as possible.
However, $\lrec$ is limited by computational constraints or by the intrinsic information content of the data: finite-width point spread functions and increasing noise at smaller angular scales limit the number of recoverable modes. These cases will be discussed in the next section. 

If $\Fmat$ is ill conditioned and $\lrec < \lgwb$, then $\hat C_{\ell}$ will be biased away from its true value, even after subtracting away $N_{\ell}$ as in \eqref{eq:Cl_estimator_1}~\cite{Thrane:2009fp,Semenzato:2025sqc}.
As shown in~\eqref{eq:noise_Bias}
\be
\llangle \hat{ C}_{\ell}\rrangle_P
=\sum_{\barl} M_{{\ell} \barl} \,C_{\barl}\,,
\label{eq:noise_Bias-main}
\ee
where~\cite{Agarwal23,Semenzato:2025sqc}
\be
M_{\ell \barl} \equiv\frac{1}{2\ell+1}\sum_{m \barm} |K_{\ell m,\barl \barm}|^2\,,
\qquad
K_{\ell m,\barl\barm} \equiv \sum_{\ell' m'} \mathcal{F}^{+}_{\ell m, \ell' m'}\, 
\mathcal{F}_{\ell' m', \barl\barm }\,.
%\sum_{a<b}\,\gamma^*_{ ab,\barl' \barm'} %\gamma_{ ab,\ell'm' }\,.
\label{eq:bias_matrix-main}
\ee
This implies that
the fractional bias in the reconstructed angular power spectrum is
\be \label{eq:bias_Cl_estimator_1}
b_{\ell} \equiv \frac{\llangle \hat{ C}_{\ell}\rrangle_P- C_{\ell}}{C_{\ell}} = \sum_{\barl} \frac{M_{{\ell} \barl}C_{\barl}}{C_{\ell}} -1\,.
\ee
Note that $\bm{M}$ and $\bm{K}$ are (nonsquare) rectangular matrices if $\ell_{\rm max}^{\rm rec}\neq \ell_{\rm max}^{\rm gwb}$. 
The behavior of the bias matrix $\bm{M}$ for a given pulsar configuration is discussed at the end of Sec.~\ref{s:leakage-bias-results:34}, where we consider {\it debiasing} the standard estimator 
\eqref{eq:Cl_estimator_1} of the angular power spectrum.

Finally, using Isserlis's theorem and the statistical properties of the estimators given in Appendix~\ref{s:stats_prop}, we can also derive the covariance matrix of the $\hat C_{\barl}$ estimators~\cite{Agarwal23}.
As shown in more detail in Appendix~\ref{s:stats_prop}, we find
\be\label{eq:Cl_cov_main}
{\rm Cov}(\hat{C}_{\ell}, \hat{C}_{\ell'})
=\frac{2}{(2\ell+1)(2\ell'+1)}\sum_{m m'} |{\cal C}_{\ell m,\ell' m'}+{\cal N}_{\ell m,\ell' m'}|^2\,,
\ee
where
\be
{\cal C}_{\ell m,\ell' m'}\equiv\sum_{\barl \barm} K_{\ell m,\barl \barm }\,K^*_{\ell' m',\barl \barm}\,C_{\barl}\,,
\label{e:Clml'm'-main}
\ee
and  ${\cal N}_{\ell m,\ell' m'}$ is given by \eqref{e:calN-main}.
We note that ${\cal C}_{\ell m,\ell' m'}$ involves a sum over $C_{\barl}$ and gives rise to \textit{cosmic variance}. In the weak-signal limit, the covariance in~\eqref{eq:Cl_cov_main} is dominated by the noise contribution ${\cal N}_{\ell m,\ell' m'}$. 

Although we have assumed the weak-signal limit to derive these expressions, the analyses presented in Sec.~\ref{s:demos} will, for the sake of simplicity, keep only the cosmic variance part of the covariance matrix (the noiseless case).
While this is not self-consistent, it is similar to the noiseless case of~\cite{Semenzato:2025sqc} against which we wish to compare our results.
(Reference~\cite{Semenzato:2025sqc} also considers the effect of pulsar noise and contributions to the variance induced by the GWB itself, which we do not.)
A more realistic treatment, including the contribution of pulsar noise, will be pursued in future work.

%%%%%%%%%%%%%%%%%%%%%%%%%%%%%%%%%%%%%%
\section{Informative angular modes in pulsar timing array data}
\label{s:lres}

The ability of a pulsar timing array to resolve angular structure in the GWB is fundamentally limited by its Fisher information. In this section, we characterize this limitation by studying properties of the Fisher information matrix, identifying an effective maximum informative multipole $\lres$.
We demonstrate its impact on angular power spectrum recovery in Sec.~\ref{s:demos}.

For these analytical calculations with the Fisher matrix, we will assume that $\sigma^2_{ab}\equiv \sigma^2\equiv 1$, corresponding to the case where the pulsar noise is the same for all pulsars.
This implies that the correlated response of the PTA to the GWB depends only the geometrical configuration of the pulsar sky locations, since it is not weighted differently according to the individual pulsar noise.

\subsection{Harmonic-space structure of the PTA Fisher matrix}
\label{s:suppression_coupling}

In the idealized limit of infinite angular resolution, the Fisher kernel would take the form
\be
\mathcal{F}(\hatom,\hatom')=\delta^2(\hatom,\hatom')\,.
\label{e:delta-lim}
\ee
(Here, we are using the word ``kernel'' when we think of 
$\mathcal{F}(\hatom,\hatom')$ as a function of two {\it continuous} directions on the sky $\hatom$ and $\hatom'$.) In this limit \eqref{e:delta-lim}, the harmonic space representation of the Fisher kernel, defined as
\be
\mathcal{F}_{\ell m, \ell' m'} \equiv \int \d^2\hatom \,\d^2 \hatom' \, Y_{\ell m} (\hatom)\,Y^*_{\ell' m'} (\hatom')\,\mathcal{F}(\hatom,\hatom')\,,
\label{e:F-harmonic}
\ee
satisfies
\be
\mathcal{F}_{\ell m, \ell' m'}=\delta_{\ell\ell'}\,\delta_{m m'}\,.
\ee 
So, all the multipoles up to $\ell_{\rm max}=\infty$ are accessible. 

For example, consider a spherically symmetric Fisher kernel derived for an infinite number of isotropically distributed
identical pulsars (a so-called ``dense-PTA''; see Eq.~(71) in~\cite{Ali-Haimoud:2020ozu}). 
By spherical symmetry, the kernel depends upon only on $\hatom\cdot \hatom'$, and its harmonic space representation is diagonal: 
\be
\label{e:gaussian_beam_harmonic}
\mathcal{F}_{\ell m, \ell' m'}=\mathcal{F}_{\ell}\,\delta_{\ell\ell'}\,\delta_{m m'}\,,
\ee
where $\mathcal{F}_{\ell}$ is a monotonically decreasing function of $\ell$ (see Fig.~3 in~\cite{Ali-Haimoud:2020ozu}). We note that: (i) because $\mathcal{F}_{\ell m, \ell' m'}$ is diagonal, there is no coupling between different multipoles, and (ii) the variance of the clean map estimator scales as $\mathcal{F}^{-1}_{\ell}$.

If we further assume statistical isotropy of the background, then there is an additional constraint: each $\ell$-mode has $2\ell+1$ independent $m$-modes that can be averaged together, thus reducing the growth in the variance for large $\ell$ values (corresponding to small angular scales). Consequently, the noise uncertainty of the angular power spectrum estimator then scales as $(2\ell+1)^{-1/2}\,\mathcal{F}^{-1}_{\ell}$, as one can see from~\eqref{eq:Cl_cov}. 

Realistic PTAs, on the other hand, have responses to the GWB that are: (i) neither spherically symmetric, (ii) nor have infinite angular resolution.
As a result of (i), the Fisher information matrix in both position and harmonic space has nonzero off-diagonal terms, corresponding to mode coupling between different multipoles.
In regard to (ii), the finite width of the position-space Fisher kernel for a realistic experiment restricts sensitivity to small-scale angular features.
The finite angular resolution blurs structures beyond a certain angular scale, acting as a low-pass filter.

\begin{figure*}[htbp!]
\centering
\includegraphics[width=0.50\textwidth]
{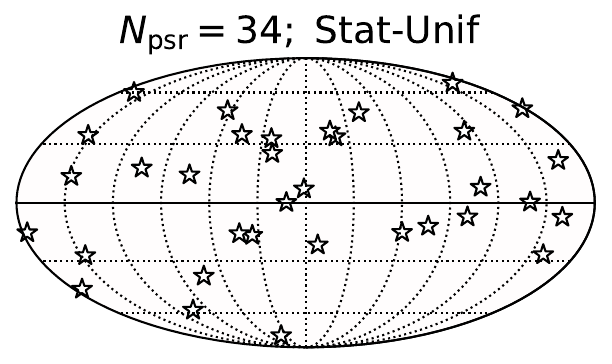}
\caption{Sky location of 34 pulsars drawn from a statistically uniform distribution, used for the main analyses in Sec.~\ref{s:demos}.}
\label{f:34pulsars}
\end{figure*}

\begin{figure*}[htb!]
\centering
\includegraphics[width=0.45\textwidth]{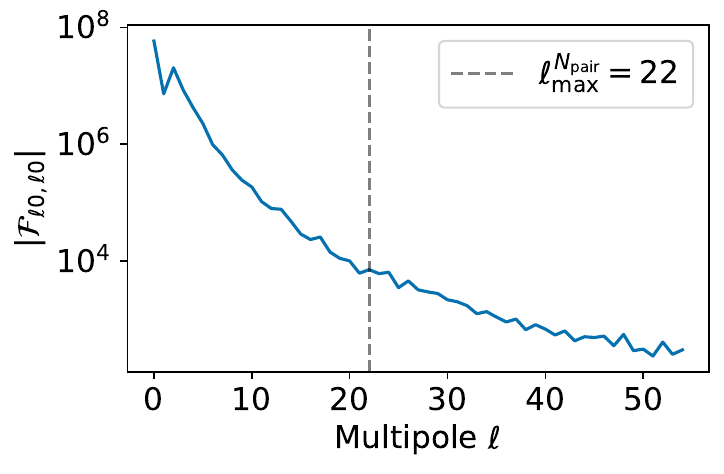}~~
\includegraphics[width=0.49\textwidth]{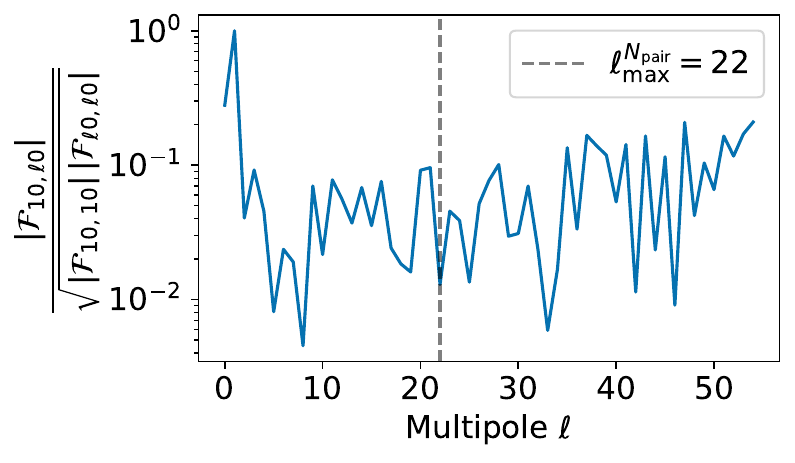}
\caption{Plots showing mode suppression (left panel) and mode coupling (right panel) calculated using the Fisher matrix for the $N_{\rm psr}=34$, Stat-Unif configuration shown in Fig.~\ref{f:34pulsars}.}
\label{f:34suppression_coupling}
\end{figure*}

To illustrate these concepts more concretely, consider a PTA configuration consisting of 34 pulsars drawn from a statistically uniform distribution on the sky, as shown in Fig.~\ref{f:34pulsars}. 
For this configuration, the diagonal elements of the Fisher information matrix decay rapidly with increasing $\ell$ (see left panel of Fig.~\ref{f:34suppression_coupling}), indicating strong mode suppression at small angular scales. 
The off-diagonal elements reveal non-negligible multipole coupling, as shown in the right panel of the same figure. 
Both the degree of mode suppression and the strength of  multipole coupling depend on specifics of the PTA configuration, as is discussed in more detail in Sec.~\ref{s:other-PTA-configs}.

Mode suppression and coupling together limit the range of angular modes that carry significant information about the distribution of GW power. This \textit{finite} scale, determined in the next section, naturally provides a truncation point for harmonic space analyses and determines the bias and leakage induced from mode coupling, as discussed in the following sections.

\subsection{Observable skies and the maximum-informative multipole $\lres$}

As mentioned in Sec.~\ref{s:intro}, $N_{\rm pair}$ cross-correlation measurements from an array of $N_{\rm psr}$ pulsars provide at most $N_{\rm pair}$ independent pieces of information (per frequency bin) about the anisotropy of the GWB~\eqref{eq:rho_ab_expectation}.
Here, we ask: how large must $\lrec$ be to capture nearly all of the significant information encoded in a PTA, while allowing for a regularization scheme to naturally suppress numerically unstable modes. 

In what follows, we will answer this question by doing the calculation using the Fisher information matrix in the PTR basis, since it captures most directly the $N_{\rm pair}$ pieces of information extracted by the pulsar pair cross-correlation measurements.
But the calculation could also be done by expressing the Fisher matrix in either the pixel or SpH bases, although for those cases, one would need to use singular-value decomposition to obtain the $N_{\rm pair}$ eigenvectors and eigenvalues. 
Provided that all $N_{\rm pair}$ eigenmodes are fully retained, the Fisher matrix in any of these bases, in principle, contain the same information.

So to answer this question about $\lres$, we start by expressing the Fisher information matrix $\Fmat$ in the PTR basis~\cite{Ali-Haimoud:2020ozu, alihaimoud2021}:
\be
\mathcal{F}_{ab,cd} \equiv \int{\rm d}^2\hatom\> 
\gamma_{ab}(\hatom)\gamma_{cd}(\hatom)
\simeq
\frac{4\pi}{N_{\rm pix}}\,\sum_{i=1}^{N_{\rm pix}}
\,\gamma_{ab}(\hatom_i)\gamma_{cd}(\hatom_i)\,,
\label{e:FIJ-int}
\ee
where the PTR maps $\gamma_{ab}(\hatom)$ are defined in \eqref{e:ORFmap}.
(Recall that we are assuming that all pulsars have the same noise, $\sigma^2_{ab}\equiv\sigma^2\equiv1$.)
The matrix component indices $ab$ and $cd$ satisfy $a<b$ and $c<d$, labeling distinct pulsar pairs.
This implies that in the PTR basis
$\mathcal{F}_{ab,cd}$ has dimension $N_{\rm pair}\times N_{\rm pair}$.
(This matrix is denoted $\mathcal{F}_{IJ}$ in \cite{Ali-Haimoud:2020ozu, alihaimoud2021}, with the indices $I=ab$, $J=cd$ labeling the distinct pulsar pairs.)

\begin{figure*}[t]%[htbp!]
    \centering
\includegraphics[width=\textwidth]{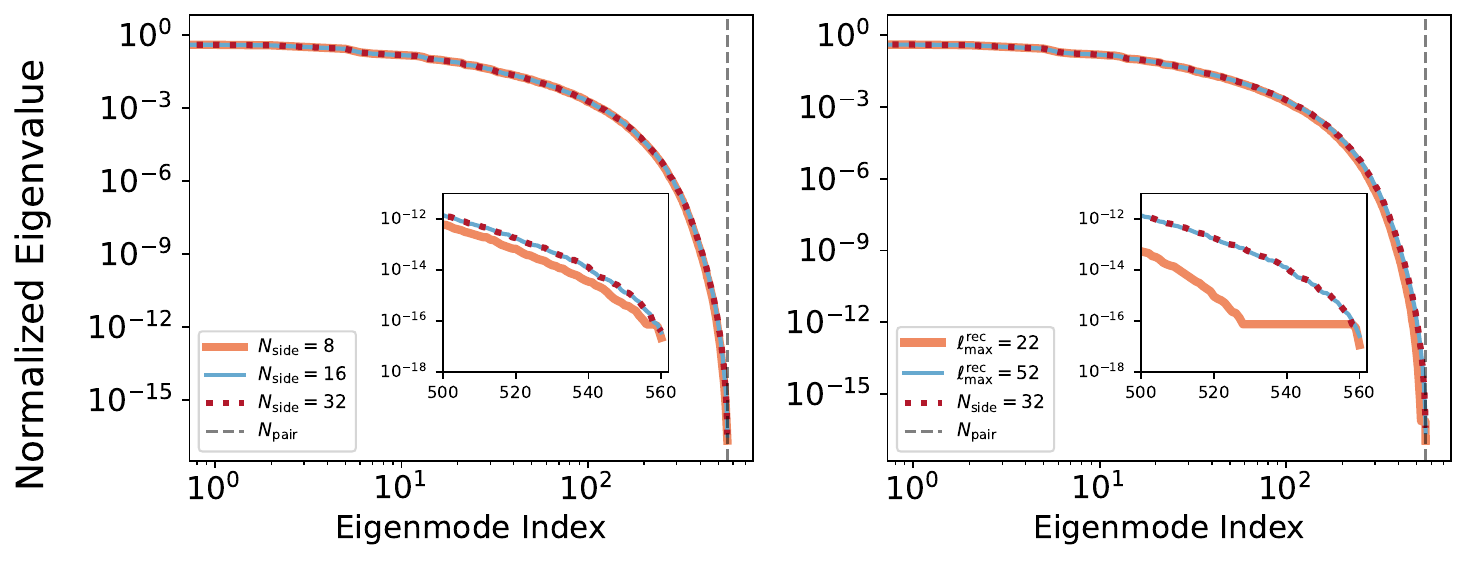}
\caption{Eigenvalue spectrum of the Fisher matrix $\mathcal{F}_{ab,cd}$ in the PTR basis, expanded in terms of pixels for different choices of $N_{\rm side}$ (left panel); and expanded in terms of SpHs for different choices of $\lrec$ (right panel). 
These plots correspond to the PTA configuration shown in Fig.~\ref{f:34pulsars}. 
Insets in both panels show zoomed-in eigenvalues for large eigenmode indices, which correspond to the smallest eigenvalues.
Note that an $N_{\rm side}=32$ pixel expansion (left panel) and an $\lrec=52$ SpH expansion (right panel) converges to the true eigenvalue spectrum, while an $N_{\rm side}=8$ pixel expansion and an $\lrec=22$ SpH expansion have not converged to the true spectrum.}
\label{f:34eigenvalue}
\end{figure*}

To evaluate the integral in \eqref{e:FIJ-int}, we need to pixelize the sky.
We choose $N_{\rm pix}$ to be sufficiently large so that the sky integral can be well approximated by a discrete sum, with results that converge as the number of pixels increase.
For all the analyses that we will describe in this paper, this requirement is satisfied if $N_{\rm pix}=12\,288$, corresponding to $N_{\rm side}=32$ in {\tt HEALPix}~\footnote{http://healpix.sf.net}; see also~\cite{2005ApJ...622..759G,Zonca2019}.  
This is shown in the left panel of Fig.~\ref{f:34eigenvalue}, 
where we compare the $N_{\rm pair}$ eigenvalues for the Fisher matrix  $\mathcal{F}_{ab,cd}$ expanded in terms of pixels for different choices for $N_{\rm side}$.
This is for the PTA configuration shown in Fig.~\ref{f:34pulsars}, which has $N_{\rm psr}=34$ pulsars and $N_{\rm pair}=561$ eigenvectors and eigenvalues.
The spectra of eigenvalues converge to the true spectrum as we increase $N_{\rm side}$ from 8 to 16 to 32.
So we will use $N_{\rm side}=32$ for all subsequent analyses in this paper, unless explicitly indicated otherwise.

We can also approximate the integral by performing a sum over the SpH components of the PTR sky maps:
\be
\mathcal{F}^{\rm sph}_{ab,cd} \equiv
\sum_{ \ell=0}^{\lrec}
\sum_{m=-\ell}^{\ell}
\gamma_{ab,\ell m}\gamma^*_{cd,\ell m}\,,
\quad{\rm where}\quad
\gamma_{ab,\ell m}\equiv
\frac{4\pi}{N_{\rm pix}}\,\sum_{i=1}^{N_{\rm pix}}
\,\gamma_{ab}(\hatom_i)Y_{\ell m}(\hatom_i)\,.
\label{e:FIJ-sph}
\ee
This sum for $\mathcal{F}^{\rm sph}_{ab,cd}$ is over $\ell$ values out to some $\lrec$ for recovery, which we recommend be $\ge\lres$, where $\lres$ captures (to sufficient accuracy, as described below) the maximum-informative 
multipole of the GWB observable by the PTA.

More precisely, we define $\lres$ as the minimum value of $\lmax$ for which $\ge 90\%$  of the $N_{\rm pair}$ eigenvectors of $\mathcal{F}^{\rm sph}_{ab,cd}$ have a match $\ge 0.95$ with the corresponding $N_{\rm pair}$ eigenvectors of the Fisher matrix $\mathcal{F}_{ab,cd}$ (expanded in terms of pixels for $N_{\rm side}=32$, as discussed above).
Here, the match between two vectors $\bm{A}$ and $\bm{B}$ is defined by
\be
{\rm Match}(\bm{A},\bm{B})\equiv
\frac{|\bm{A}\cdot\bm{B}|}{\sqrt{(\bm{A}\cdot\bm{A})(\bm{B}\cdot\bm{B})}}\,,
\ee
where $\bm{A}\cdot\bm{B}$ denotes the usual Euclidean inner product of two vectors, which is the sum of the product of their components with respect to an orthonormal basis. 
Note that we have included an absolute value sign in the definition of the match to allow for a possible $\pm$ ambiguity in the definition of the eigenvectors.

\begin{figure*}[t]%[htbp!]
    \centering
\includegraphics[width=0.5\textwidth]{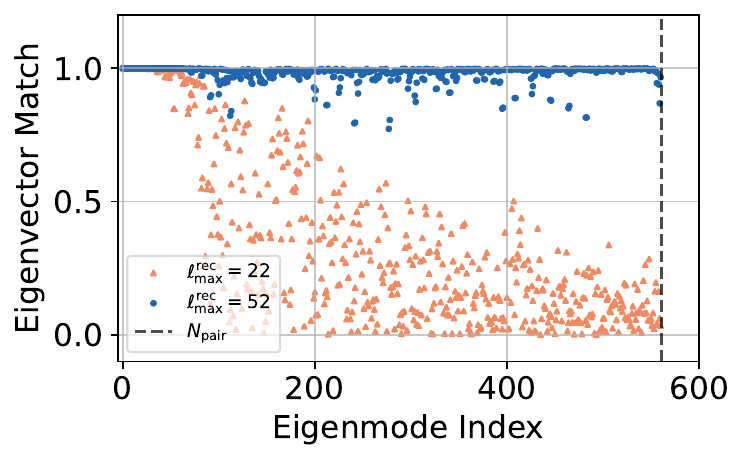}~~
\includegraphics[width=0.5\textwidth]{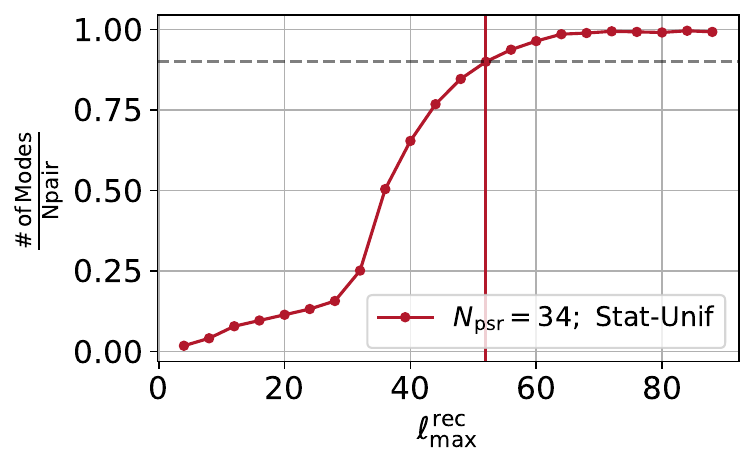}
\caption{Left panel: match between eigenvectors of the Fisher matrix $\mathcal{F}_{ab,cd}$ in the PTR basis, expanded in terms of SpHs for different choices of $\lrec$.
(The expansion of the eigenvectors of $\mathcal{F}_{ab,cd}$ in terms of pixels having $N_{\rm side}=32$ is used as reference.)
The SpH expansion with $\ell_{\rm max}^{\rm rec}=52$ (blue circles) recovers most ($>90\%$) of the $N_{\rm pair}$ eigenvectors (vertical dashed), while $\ell_{\rm max}^{\rm rec}=22$ (orange triangles) captures only the lowest modes. 
Right panel: fraction of eigenmodes with match in the range (0.95,1) normalized by $N_{\rm pair}$, as a function of $\lrec$. The vertical line marks $\ell_{\rm max}^{\rm res}=52$, the maximum informative angular scale for this configuration corresponding to fraction 0.9 (horizontal dashed gray line).
Both of these plots correspond to the PTA configuration shown in Fig.~\ref{f:34pulsars}.}
\label{f:34eigenvector}
\end{figure*}

The result of this match calculation leads to $\lres=52$ for the $N_{\rm psr}=34$ PTA configuration shown in Fig.~\ref{f:34pulsars}.  
Details of this calculation are illustrated in  Fig.~\ref{f:34eigenvector}, including a comparison (in the left panel) of two different analyses: one which chooses $\lrec=\lres=52$ (the maximum informative multipole for the array), while the other chooses $\lrec=\lNpair=22$ (the conventional limit for $34$ pulsars).
By choosing $\lrec=52$ (blue circles), we obtain very good approximations of the Fisher matrix eigenvectors, thus allowing us to recover the majority of the $N_{\rm pair}$ informative modes.
Choosing $\lrec=22$ (orange triangles), on the other hand, leads to poor approximations of these eigenvectors. 
In the right panel of Fig.~\ref{f:34eigenvector}, we plot the fraction of modes with match $\ge 0.95$ as a function of $\lrec$.
The curve exceeds 0.90 when $\lrec=52$ (vertical red line), identifying $\lres=52$ as the maximum informative angular scale for the PTA configuration in 
Fig.~\ref{f:34pulsars}. 
In addition, as shown in the right panel of Fig.~\ref{f:34eigenvalue}, $\lrec=52$ (blue solid) consistently recovers the true eigenvalue spectrum (red dashed), whereas $\lrec=22$ (orange solid) does not.

%%%%%%%%%%%%%%%%%%%%%%%%%%%%%%%%%%%%%%%%%%%%%
\section{Demonstrations of leakage and mode suppression}
\label{s:demos}

In this section, we illustrate the impact of mode suppression and mode coupling, and compare the reconstruction of the angular power spectrum with a PTA when analyses are truncated at $\lrec$ equal to either $\lNpair$ or $\lres$.
For the following analyses we will assume that the pulsar noise is zero, i.e., $\sigma^2_{ab}=0$ for all pulsar pairs.
As mentioned earlier, this allows us to easily compare our results to \cite{Semenzato:2025sqc}, which also considers leakage of small-scale angular power in this context (\cite{Semenzato:2025sqc} also considers noise-full analyses).

We note that setting $\sigma^2_{ab}=0$ is well defined and does not lead to ill-defined behavior in the equations. This is because the clean map estimator~\eqref{eq:clean_estimator} is independent of $\sigma^2$ once the same noise-level limit is imposed.
In this limit, its noise variance vanishes as the Fisher information matrix~\eqref{eq:dirty_fisher} diverges. 
As a result, the signal contribution to both the standard $C_\ell$ estimator~\eqref{eq:Cl_estimator_1} and its covariance~\eqref{eq:Cl_cov_main} remain unchanged, while the noise contribution to the bias (equal to the clean map noise variance) and covariance are identically zero. 
For this reason, we refer to this noise-free case as the best-case scenario for angular power spectrum estimation. 
In the presence of (unknown) noise, obtaining an unbiased estimator becomes more challenging---particularly due to limitations on the reconstructable modes---but this regime is beyond the scope of the present work.

\subsection{Mode-coupling-induced leakage and $\lres$}
\label{s:leakage-bias-results:34}

\begin{figure*}[h!]
\centering
    \includegraphics[width=\textwidth]{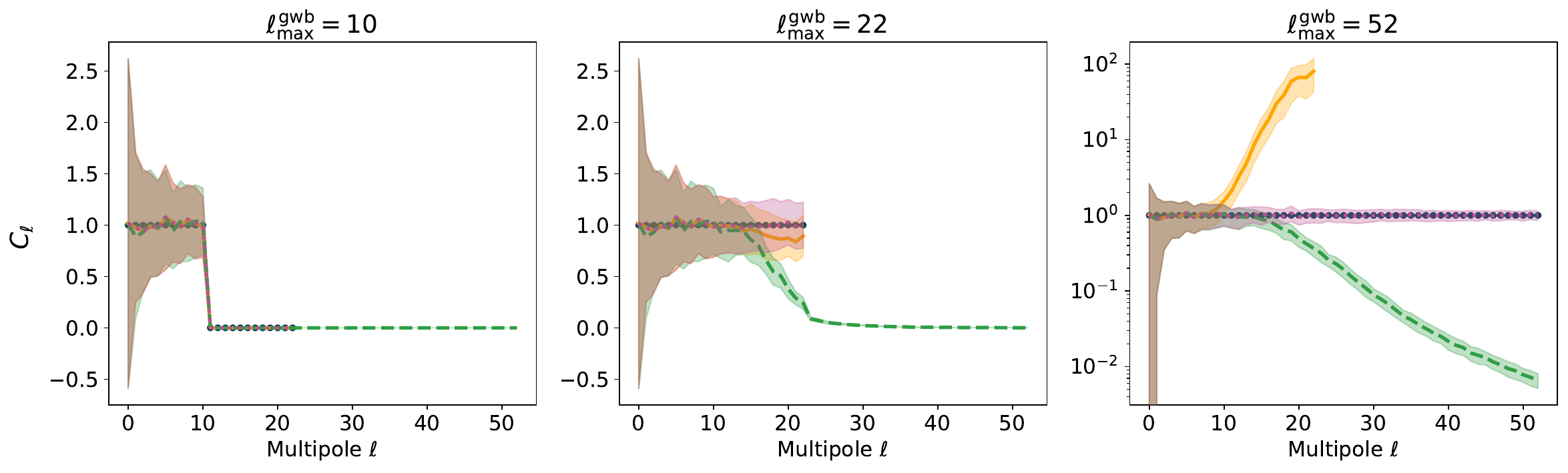}
    \includegraphics[width=0.6\textwidth]{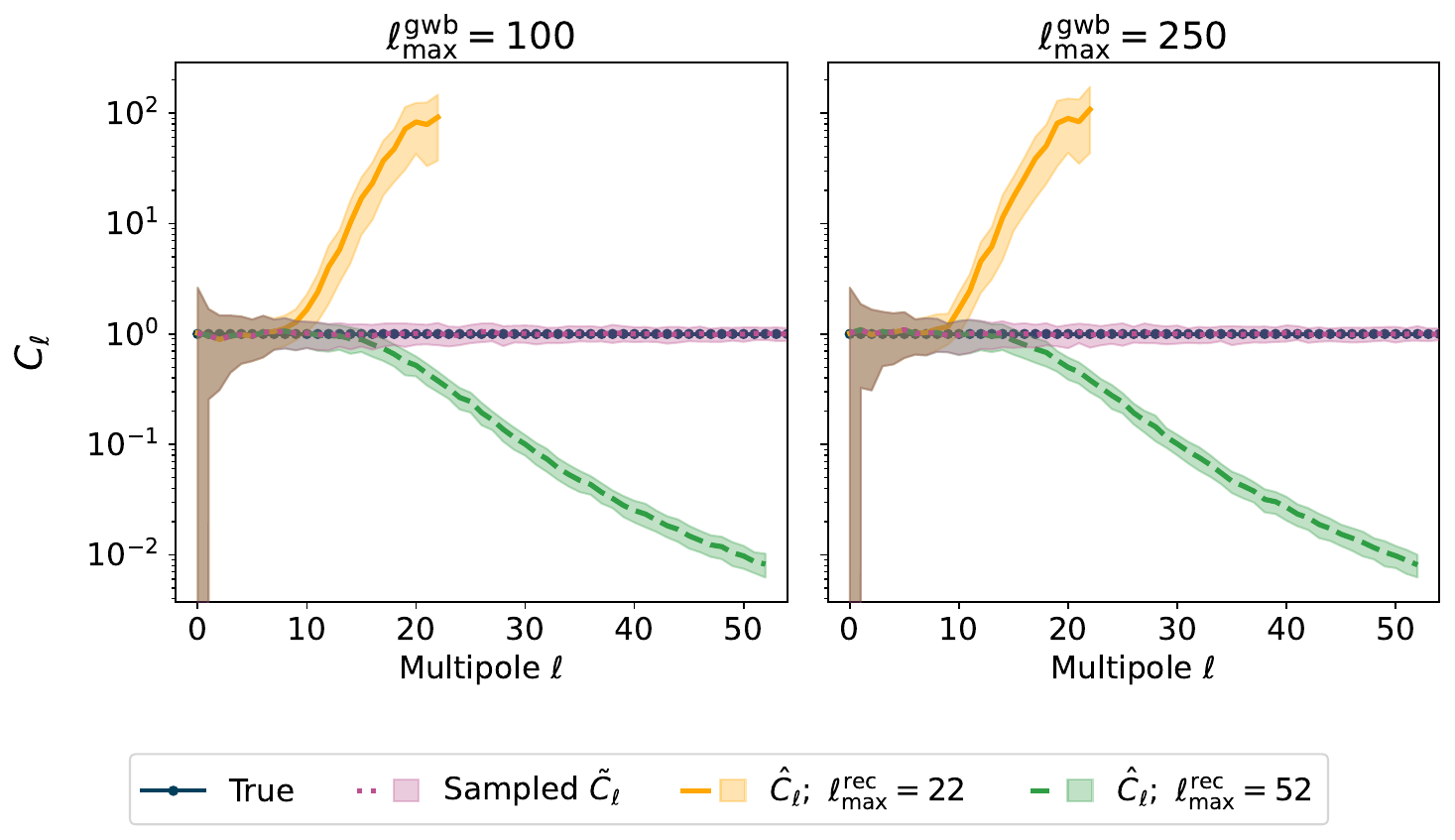}
    \caption{Demonstration of leakage-induced bias in the reconstructed angular power spectrum $\hat C_\ell$ for the PTA configuration shown in Fig.~\ref{f:34pulsars}. As $\lgwb$ increases from 10 to 52 (left to right panels in top row), a significant positive bias appears when truncating recovery at $\ell_{\rm max}^{\rm rec}=22$ (yellow solid), but not when using $\ell_{\rm max}^{\rm rec}=52$ (green dashed). The reconstruction is compared with the true (theoretical) $C_\ell$ (blue solid-circle) and sampled-universe $\tilde C_\ell$ (pink dotted). The shaded region around the estimated $\hat C_\ell$ values denotes $\pm 1\sigma$ cosmic uncertainty.
    Increasing $\lgwb$ beyond 52 does not significantly affect the reconstruction (within cosmic variance; bottom row), indicating that power at higher multipoles does not propagate through the array.}
    \label{f:34Clrecovery_top}
\end{figure*}

As discussed in the previous section, $\lNpair$ does not span the full observable space; therefore setting $\lrec=\lNpair$ leads to a loss of important information.
To capture most of the informative modes, one should instead adopt $\lrec\ge\lres$.
To demonstrate these claims, we perform numerical simulations for constructing the angular power spectrum where $\lrec$ is chosen to equal either $\lNpair$ or $\lres$. 
The key results from this simulation study are shown in Fig.~\ref{f:34Clrecovery_top}. The simulations are constructed as follows:
\begin{itemize}
\item We consider a PTA configuration consisting of 34 pulsars drawn from a statistically uniform distribution on the sky (see Fig.~\ref{f:34pulsars}).
    
\item No pulsar noise is injected, corresponding to an idealized best-case scenario.

\item We inject a flat angular power spectrum, $C_\barl=1$ for $0\leq\barl\leq \lgwb$, with $\lgwb$ taking values $\{10,22,52,100,250\}$.

\end{itemize}
The blue solid line with circle markers in Fig.~\ref{f:34Clrecovery_top} represent the true angular power spectrum, $C_\barl$. For each value of $\lgwb$, we generate 100 realizations of the SpH coefficients $P_{\barl\barm}$ assuming statistical isotropy~\eqref{eq:stats_Iso}. For each realization, the angular power spectrum is estimated as 
\be
\tilde{C}_\barl=\frac{1}{2\barl+1}\sum_m |P_{\barl\barm}|^2\,.
\ee 
In Fig.~\ref{f:34Clrecovery_top}, the pink dotted line and shaded band show the mean and $\pm1\sigma$ scatter of $\tilde C_\barl$ over the simulated realizations, respectively, corresponding to the cosmic mean and cosmic variance.

Using these realizations, we simulate the response of the pulsars to the GWB and construct clean map estimators, $\hat{P}_{\ell m}$ following the procedure described in Sec.~\ref{s:preliminaries}. The Fisher matrix is regularized by discarding modes with condition number threshold $\kappa=10^{13}$. 
The reconstructed angular power spectrum $\hat{C}_\ell$ is then obtained using~\eqref{eq:Cl_estimator_1}.

For each value of $\lgwb$, we compare the reconstructed angular power spectrum for two choices of $\lrec$: 
(i) $\lrec=\lNpair=22$ (yellow solid), and (ii) $\lrec=\lres=52$ (green dashed). 
We observe the following:
\begin{itemize}
    
\item For $\lgwb=10$, the reconstruction is unbiased for both $\lrec=22$ and $\lrec=52$ (top-left panel of Fig.~\ref{f:34Clrecovery_top}).
    
\item For $\lgwb=22$, the reconstruction is slightly negative-biased for $\lrec=22$, while a significant negative bias appears for $\lrec=52$ (top-middle panel of Fig.~\ref{f:34Clrecovery_top}).

\item For $\lgwb=52$, a positive bias appears for $\lrec=22$, corresponding to the leakage reported in~\cite{Semenzato:2025sqc} (top-right panel of Fig.~\ref{f:34Clrecovery_top}). 
There is no positive bias for $\lrec=52$; 
the negative bias for $\lrec=52$ seen previously for $\lgwb=22$ persists.

\item For larger values of $\lgwb$, namely $\lgwb =100, \>250$ (bottom two panels in Fig.~\ref{f:34Clrecovery_top}), the positive bias persists for $\lrec=22$ with no significant dependence on $\lgwb$ (an enlarged plot would show the change is within the uncertainty associated with cosmic variance).
This indicates that negligible additional information beyond $\lres=52$ makes its way into the angular power spectrum estimates out to $\lrec=22$ and $\lrec=52$.
\end{itemize}

These results confirm that the positive bias reported in~\cite{Semenzato:2025sqc} when reconstructing the angular power spectrum  originates from unmodeled informative modes when the reconstruction is truncated at $\lrec=22$. 
In \cite{Semenzato:2025sqc}, the angular scales are divided into ``large'' and ``small'' scales using a cutoff defined by $\lNpair$, which is determined solely by the number of pulsars, without taking into account their sky distribution. 
Consequently, informative modes with $\lNpair<\ell\le \lres$ are omitted from their reconstruction, but they couple to the retained modes, producing the observed leakage bias.

Our analysis demonstrates that this leakage is not a fundamental consequence of PTA observations, but rather an artifact of truncating the reconstruction without adequately representing that observable subspace spanned by the pulsar array. 
Using the Fisher information matrix, we have identified a physically motivated reconstruction scale, $\lres$, which depends on specifics of the pulsar configuration.
We find that the leakage is primarily due to modes having $\lNpair<\ell\le \lres$. 
Choosing $\lrec=52$ incorporates these informative modes and effectively removes the leakage-induced bias. 
An additional advantage of adopting $\lrec=52$ is that, for GWBs having $\lgwb =52$ and higher, we are able to recover approximately seven additional $C_\ell$ values in the range $\ell\in[11,17]$, consistent with their injected values to within $\pm1\sigma$ uncertainties, in contrast to the case with $\lrec=22$ (top right panel of Fig.~\ref{f:34Clrecovery_top}).

We also note the observed leakage is largely insensitive to the choice of regularization.
This result was observed in~\cite{Semenzato:2025sqc} and is confirmed by our analysis, which uses a different regularization scheme from that in~\cite{Semenzato:2025sqc}.
The fact that we reproduce the same leakage-induced bias using a different regularization scheme supports the conclusion that the bias originates from omission of the informative modes in the reconstruction and is not due to some issue associated with regularization.

We note that for $\lrec=52$ a negative bias persists in all cases for GW modes beyond $\ell\approx 18$. This bias reflects the partial (and eventually complete) loss of information due to the finite angular resolution of the PTA configuration, as discussed in detail in Sec~\ref{s:suppression_coupling}. 
In the next subsection, we investigate whether it is possible to recover some of this partially lost information by constructing an unbiased estimator for the angular power spectrum.

\begin{figure*}[htb!]
\centering
\includegraphics[width=\textwidth]{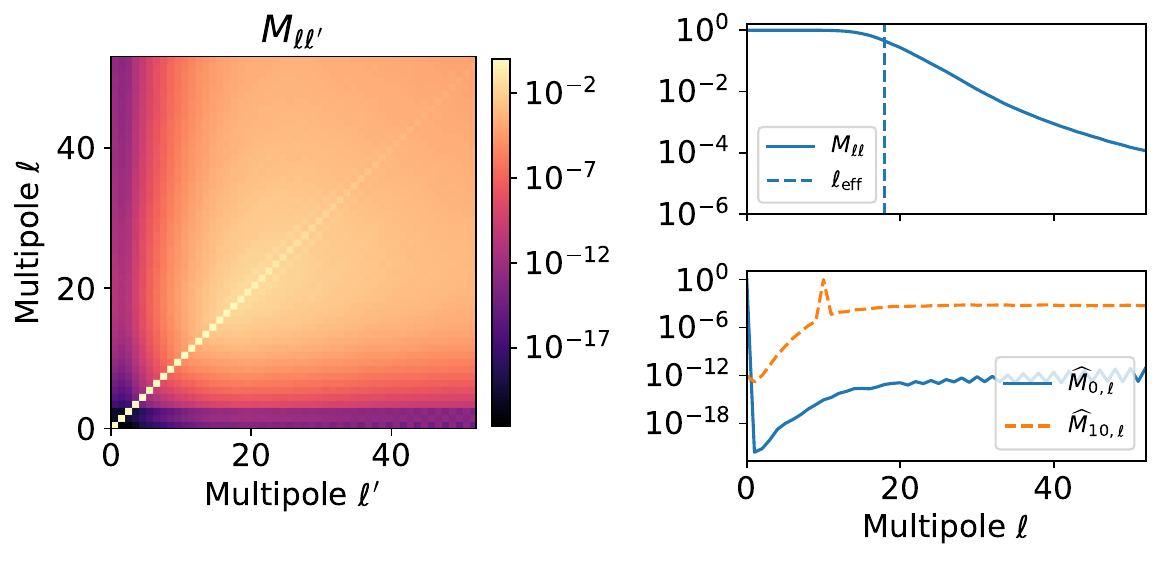}
\caption{Left panel: Bias matrix $\bm{M}$~\eqref{eq:bias_matrix-main} for the pulsar configuration shown in Fig.~\ref{f:34pulsars}, with $\lrec=\lgwb=52$ and $\kappa=10^{13}$. Top right panel: diagonal elements of the matrix, with the vertical line marking the effective multipole $\leff$ (defined as the multipole where $M_{\ell \ell}=M_{00}/2$). Bottom right panel: mode coupling in the normalized bias matrix, $\widehat{M}_{\ell \ell'}\equiv M_{\ell \ell'}/\sqrt{M_{\ell \ell}\,M_{\ell' \ell'}}$.}
\label{f:34Mmatrix}
\end{figure*}

Based on the analytical bias matrix $\bm{M}$~\eqref{eq:bias_matrix-main}, we define an effective maximum multipole $\leff$, corresponding to the angular scale up to which the angular power spectrum can be reliably recovered. 
We identify this scale as the multipole at which the diagonal element $M_{\ell\ell}$ drops to half of its maximum value, $M_{00}$. 
Since $\bm{M}$ depends upon the pulsar configuration and the condition number threshold $\kappa$ used in calculating the pseudoinverse of the Fisher matrix $\bm{\mathcal{F}}$, the effective multipole $\leff$ is likewise dependent on these choices. 
For the pulsar array shown in Fig.~\ref{f:34pulsars} and $\kappa=10^{13}$, the analytical bias matrix is shown in the left panel of Fig.~\ref{f:34Mmatrix}, with the effective multipole estimated as $\leff\approx 18$ (indicated by vertical line in the top right panel of Fig.~\ref{f:34Mmatrix}). 
We note that the multipoles are approximately uncorrelated at low $\ell$, while significant mode coupling exists at higher multipoles, as shown in the bottom-right panel of Fig.~\ref{f:34Mmatrix}.
Finally, the analytical bias expression~\eqref{eq:bias_Cl_estimator_1} shows that, in the absence of mode coupling, no bias is introduced even when $\lrec<\lres$ as demonstrated in Appendix~\ref{app:noBiasNoCoupling}.

\subsection{Debiasing the standard $C_{\ell}$ estimator}

Imposing the requirement that the angular power spectrum estimator be unbiased, i.e.,
\be
\llangle\hat{C}^{(u)}_{\ell}\rrangle_P = C_{\ell}\,,
\label{e:Chat_ub}
\ee
we define an unbiased estimator
\be
\label{e:debiased_estimator_main}
 \hat{C}_{\ell}^{(u)} \equiv \sum_{\ell'}\,(M^{-1})_{\ell\ell'}\,\hat{C}_{\ell'} \,,
\ee
where $\hat{C}_{\ell}$ and the bias matrix $\bm{M}$ are given by~\eqref{eq:Cl_estimator_1} and~\eqref{eq:bias_matrix-main}, respectively; see also~\eqref{eq:noise_Bias-main}.
(Note that we have performed debiasing only in the recovered multipole space, summing over $\ell'$ out to $\lrec$.
In other words, we are assuming that we have no prior knowledge of the injected power spectrum for multipoles greater than $\lrec$.)

But it is important to note that the bias matrix $\bm{M}$ may be ill conditioned (similar to what we saw for the Fisher information matrix $\bm{\Fmat}$), necessitating the use of a pseudoinverse $\bm{M}^+$ for $\bm{M}^{-1}$.
This implies that \eqref{e:debiased_estimator_main} should be replaced by 
\be
\label{e:debiased_estimator_regularized_main}
 \hat{C}_{\ell}^{(u)} \equiv \sum_{\ell'}\,M^+_{\ell\ell'}\,\hat{C}_{\ell'} \,,
\ee
which now depends on how we choose to regularize $\bm{M}$.
For calculating the pseudoinverse, we adopt the regularization scheme described in Eqs.~\eqref{e:SVD_1}--\eqref{e:SVD_3}. 
Note that this regularization introduces a bias-variance trade off, similar to that discussed earlier in the context of regularizing the Fisher matrix.
This means that the resulting estimator depends upon the chosen condition number threshold.
The impact of the choice of condition number threshold on $C_\ell$ recovery for the debiased estimator is discussed in Appendix~\ref{s:impact_debiased}.

Using the pseudoinverse of the bias matrix $\bm{M}$ in \eqref{e:debiased_estimator_regularized_main} also means that the debiased estimator $\hat C^{(u)}_\ell$ is not actually unbiased.
It still has a residual bias, which we see when calculating it expected value
\be
\label{e:debiased_mean_main}
\llangle \hat C^{(u)}_{\ell}\rrangle_P
=\sum_{\ell'}\sum_{\barl}
M^+_{\ell\ell'}
M^{\phantom{+}}_{\ell'\barl}C_{\barl}\,.
\ee
The rhs of the above expression does not equal $C_\ell$ since $\bm{M}^+\bm{M}$ is not the identity matrix. 
For completeness, it is also relatively straightforward to write down the covariance matrix of the debiased estimators $\hat C^{(u)}_{\ell}$ and $\hat C^{(u)}_{\ell'}$:
\be
\label{eq:debiased_cov_main}
\ba
{\rm Cov}(\hat C^{(u)}_{\ell}, \hat C^{(u)}_{\ell'})
&= \sum_{\ell_1}\sum_{\ell_2}M^+_{\ell \ell_1}
{\rm Cov}(\hat C^{(u)}_{\ell_1}, \hat C^{(u)}_{\ell_2})M^+_{\ell_2, \ell'}
\\
&=
\sum_{\ell_1}\sum_{\ell_2}\frac{2}{(2\ell_1+1)(2\ell_2+1)}\sum_{mm'} M^+_{\ell \ell_1}|{\cal C}_{\ell_1 m,\ell_2 m'}+{\cal N}_{\ell_1 m,\ell_2 m'}|^2 M^+_{\ell_2, \ell'}\,,
\ea
\ee
which follows immediately from \eqref{e:debiased_estimator_regularized_main} and the  covariance \eqref{eq:Cl_cov_main} of the standard estimator $\hat C_\ell$.

\begin{figure*}[h!]
    \centering
    \includegraphics[width=\linewidth]{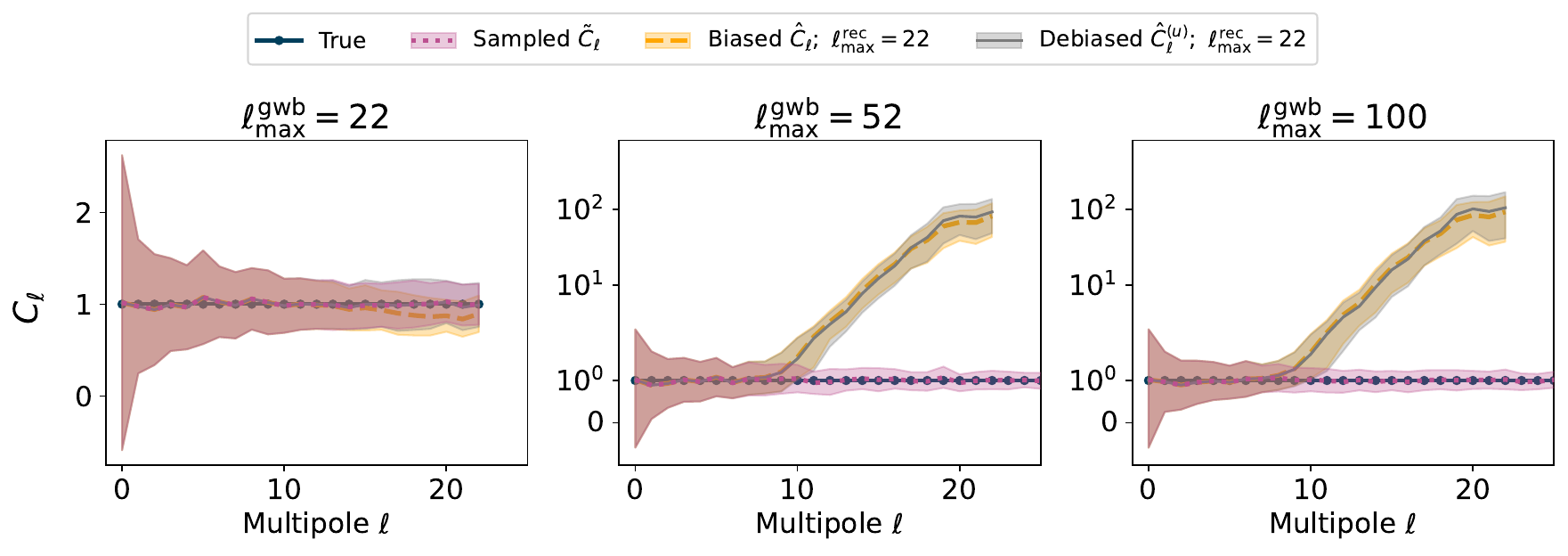}
    \caption{Debiased angular power spectrum recovery for $\lrec=22$. Comparison of the standard estimator $\hat{C}_\ell$ (yellow dashed), the debiased estimator $\hat{C}^{(u)}_\ell$ (black solid), and the injected spectrum (blue solid-circle). The simulation setup is identical to that shown in Fig.~\ref{f:34Clrecovery_top}. Debiasing corrects the recovery for the $\lgwb=22$ case, but does not significantly modify the results for $\lgwb=52$ and $\lgwb=100$, and the leakage-induced bias persists for multipoles affected by unmolded informative angular scales---i.e., $\lrec<\ell\le\lres$.}
    \label{fig:debiased_22}
\end{figure*}

We apply this debiasing procedure to the simulations described in Sec.~\ref{s:leakage-bias-results:34}. 
We first consider the case $\lrec=22$, with results shown in Fig.~\ref{fig:debiased_22}. 
For this case, the condition number of the bias matrix $\bm{M}$ matrix is $\sim\!1.3$, so no regularization is required. 
This debiasing procedure is able to correct for the negative bias arising for the case $\lgwb=22$ (left panel of Fig.~\ref{fig:debiased_22}), but not for the cases where $\lgwb>\lrec$. 
This indicates that, in the absence of additional spectral information or forward modeling, debiasing alone cannot fully remove the leakage-induced bias.
\begin{figure*}[h!]
    \centering    \includegraphics[width=\linewidth]{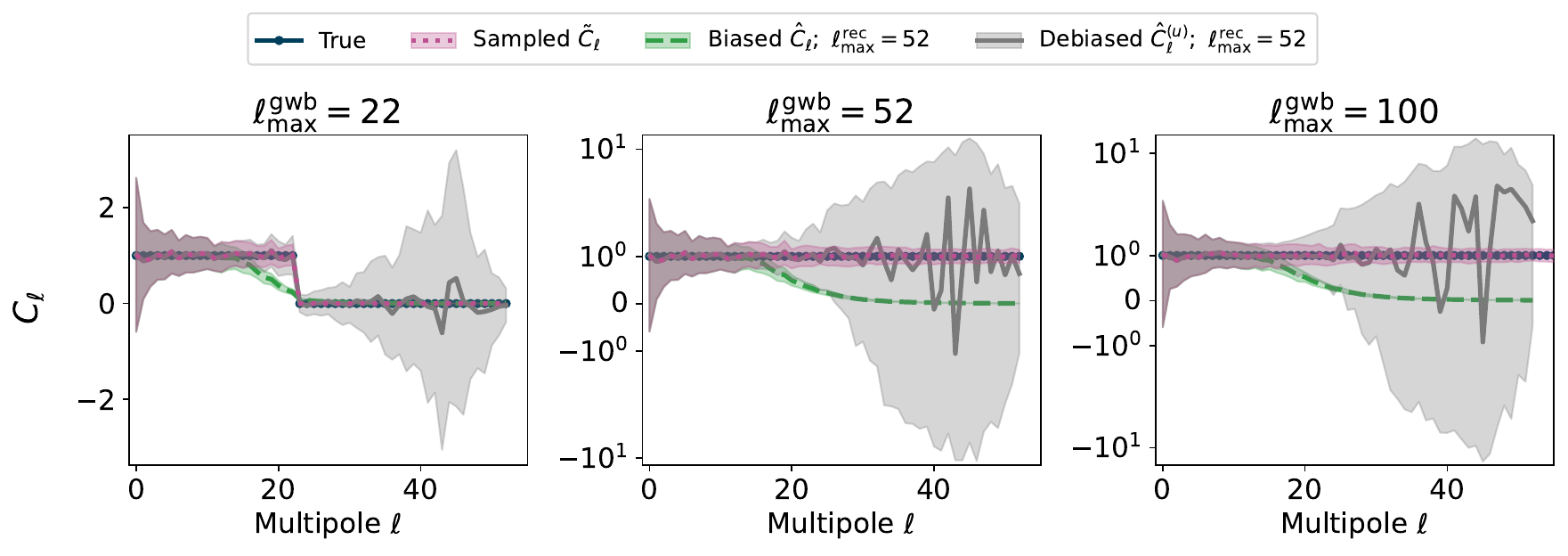}
    \caption{Similar to Fig.~\ref{fig:debiased_22}, but for $\lrec=52$, for simulations shown in Fig.~\ref{f:34Clrecovery_top}.
    Mean recovered $C_\ell$ with $\pm 1\sigma$ uncertainties (shaded bands) compared to the injected spectrum (blue solid circle). 
    Results are shown for condition number threshold $=10^4$ used to construct the pseudoinverse for the bias matrix $\bm{M}$. 
    Debiasing improves agreement with the injected spectrum for well-resolved multipoles ($\lesssim 22$), while the variance increases for poorly constrained modes ($>22$). 
    Figure~\ref{fig:debiased_52_vs_condition} in Appendix~\ref{s:impact_debiased} illustrates how the choice of condition number threshold impacts the recovery of the angular power spectrum components $C_\ell$.}
    \label{fig:debiased_52}
\end{figure*}

We next perform debiasing for the case of $\lrec=52$.
For this case, $\bm{M}$ is not invertible and requires regularization to compute its pseudoinverse.
The recoveries shown in Fig.~\ref{fig:debiased_52} correspond to a condition number threshold $=10^4$.
Compared to the recovery of the $C_\ell$'s using the standard estimator, debiasing improves agreement with the injected spectrum for well-resolved multipoles ($\lesssim 22$).
The variance increases for $\ell >22$ corresponding to poorly constrained modes. 
Results for recoveries of the angular power spectrum for  $\lrec=52$ for different choices of the condition number threshold are discussed in Appendix~\ref{s:impact_debiased}.

We do not show results for $\lgwb=10$ and 250 in Figs.~\ref{fig:debiased_22} and \ref{fig:debiased_52} for the following two reasons:
(i) for $\lgwb=10$, the standard estimator already yields unbiased recovery (left panel of Fig.~\ref{f:34Clrecovery_top}), and debiasing has no effect; and 
(ii) for $\lgwb=250$, the recovered spectrum is consistent within cosmic uncertainty with the cases $\lgwb=52$ and 100, as shown in Fig.~\ref{f:34Clrecovery_top}.

\begin{figure*}[htbp!]
    \centering   
    \includegraphics[width=0.4\linewidth]{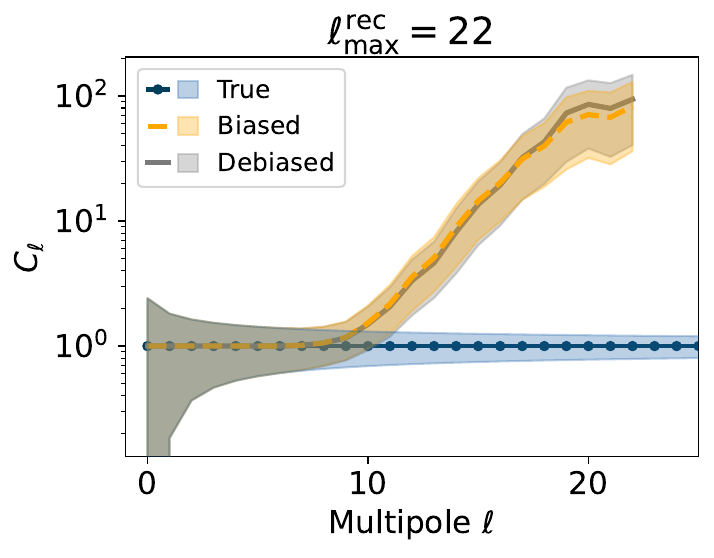}
    \includegraphics[width=0.4\linewidth]{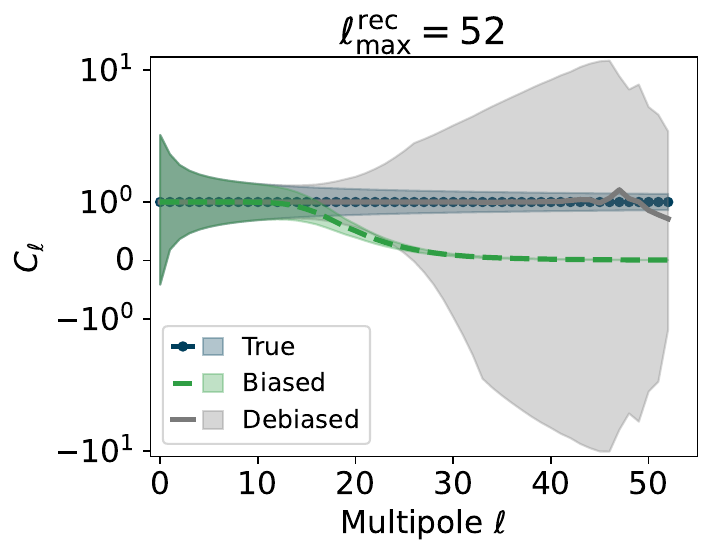}
    \caption{Similar to the middle panels of Figs.~\ref{fig:debiased_22} and ~\ref{fig:debiased_52}, but using analytical expressions for the expected value of the debiased estimator $\hat C^{(u)}_\ell$  of the angular power spectrum $\pm$ its $1\sigma$ uncertainty using \eqref{e:debiased_mean_main} and \eqref{eq:debiased_cov_main}.
    For these plots, $\lgwb=52$.}
    \label{f:debiased_analytic}
\end{figure*}

Finally, we note that the results of the numerical simulations shown in Figs.~\ref{fig:debiased_22} and \ref{fig:debiased_52}  for the 
debiased estimator $\hat C^{(u)}_{\ell}$ 
are consistent with analytical expressions for the expected value and variance of this estimator,  given in \eqref{e:debiased_mean_main} and \eqref{eq:debiased_cov_main}.
Figure~\ref{f:debiased_analytic} shows $\llangle \hat C^{(u)}_\ell\rrangle\pm \sigma_\ell$, calculated using \eqref{e:debiased_mean_main} and \eqref{eq:debiased_cov_main} for both $\lrec=22$ (left panel) and $\lrec=52$ (right panel), for the case $\lgwb=52$.
The condition number threshold for $\bm{M}$ used for the $\lrec=52$ plot is $10^4$, which agrees with that used for the plots in Fig.~\ref{fig:debiased_52}.
One should compare the left and right panels in Fig.~\ref{f:debiased_analytic} with the middle panels in Figs.~\ref{fig:debiased_22} and \ref{fig:debiased_52}, respectively. 
The agreement between these plots is a sanity check on the correctness of the analytical expressions for the mean and variance of the debiased estimator.

Thus, performing angular power spectrum estimation choosing $\lrec=\lres$, together with an optimal choice of the condition number threshold for inverting the Fisher information and bias matrices, leads to an effectively unbiased and leakage-free recovery of the angular power spectrum, in contrast to simply truncating the SpH analysis at $\lNpair$.

%%%%%%%%%%%%%%%%%%%%%%%
\section{Dependence on PTA configuration}
\label{s:other-PTA-configs}

In the previous section, we showed that the space of ``observable skies" for a PTA (i.e., the eigenspace of the corresponding Fisher matrix) is spanned by angular scales up to $\lres$,
which is generally larger than the theoretical limit $\lNpair$ set by the number of pulsar pairs. 
In this section, we investigate the dependence of $\lres$ on details of the pulsar configuration---specifically, how $\lres$ changes when the number of pulsars is held fixed while their sky distribution varies, or when the number of pulsars is increased while preserving the statistical properties of their spatial distribution.

\begin{figure*}[htb!]
\centering
\includegraphics[width=0.49\textwidth]{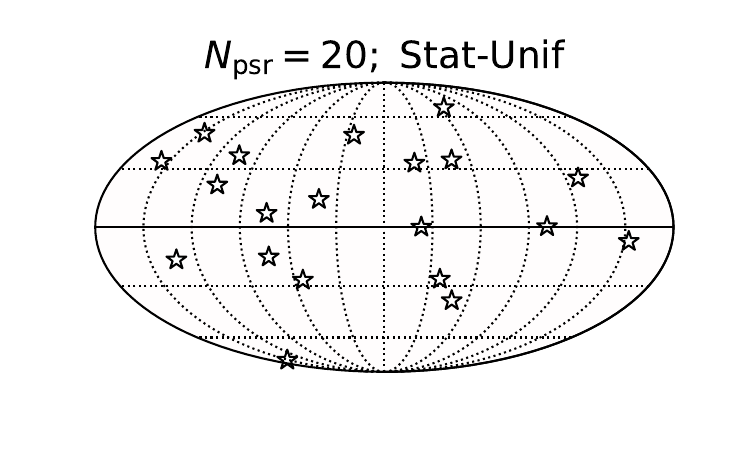}
\includegraphics[width=0.49\textwidth]{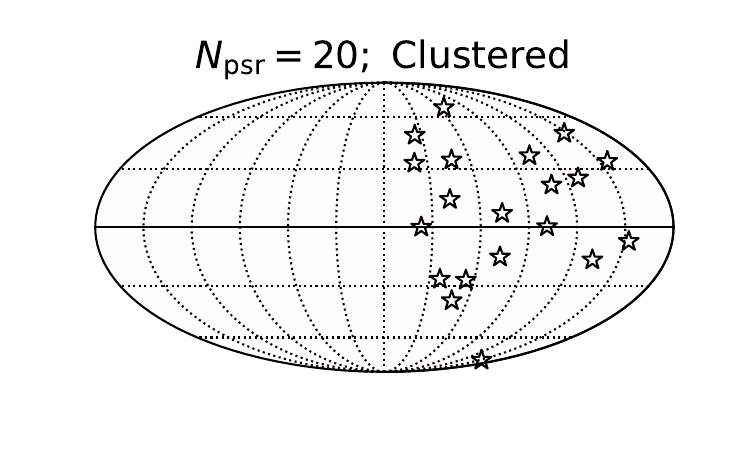}
\includegraphics[width=0.49\textwidth]{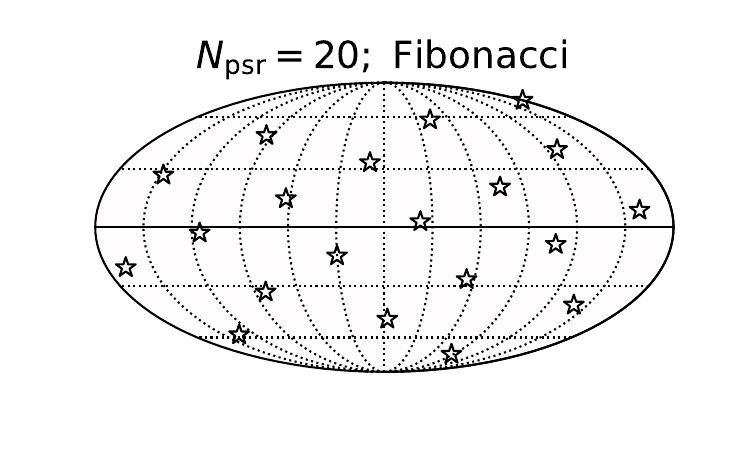}
\includegraphics[width=0.49\textwidth]{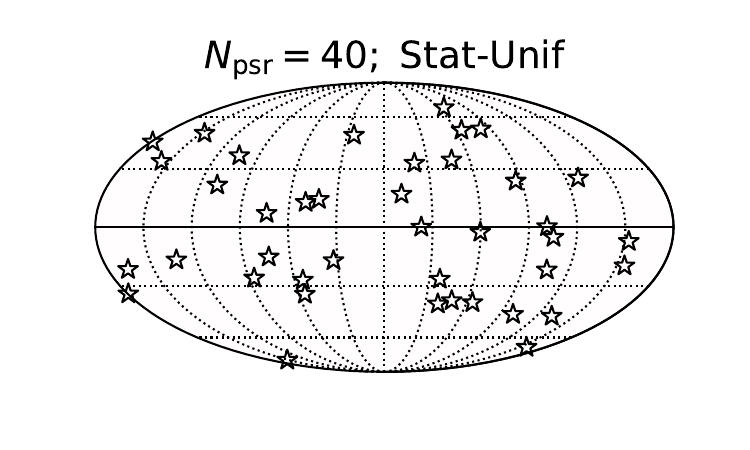}
\caption{Sky locations of 20 and 40 pulsars for the four different PTA configurations used for the analyses in Sec.~\ref{s:other-PTA-configs}.
The $N_{\rm psr}=20$, Clustered configuration is obtained by reflecting the 11 pulsars in the ``left hemisphere'' of $N_{\rm psr}=20$, Stat-Unif configuration to the ``right hemisphere''.
The $N_{\rm psr}=40$, Stat-Unif configuration is obtained by supplementing the $N_{\rm psr}=20$, Stat-Unif configuration with 20 additional pulsars drawn from a  statistically uniform distribution on the sky.
The $N_{\rm psr}=20$, Fibonacci configuration distributes the pulsars on a Fibonacci lattice, yielding a more uniform separation of neighboring pulsars.}
\label{f:20pulsars+}
\end{figure*}

To facilitate this investigation, we consider four PTA configurations shown in Fig.~\ref{f:20pulsars+}:
\begin{enumerate}
    \item $N_{\rm psr}=20$, Stat-Unif: 20 pulsars drawn from a statistically uniform distribution on the sky.
    \item $N_{\rm psr}=20$, Clustered: 20 pulsars concentrated in a single hemisphere.
    \item $N_{\rm psr}=20$, Fibonacci: 20 pulsars distributed according to a Fibonacci lattice, yielding a more uniform separation of neighboring pulsars on the sky.
    \item $N_{\rm psr}=40$, Stat-Unif: 40 pulsars drawn from a statistically uniform distribution on the sky.
\end{enumerate}
\begin{figure*}[htb!]
\centering
\includegraphics[width=0.75\textwidth]{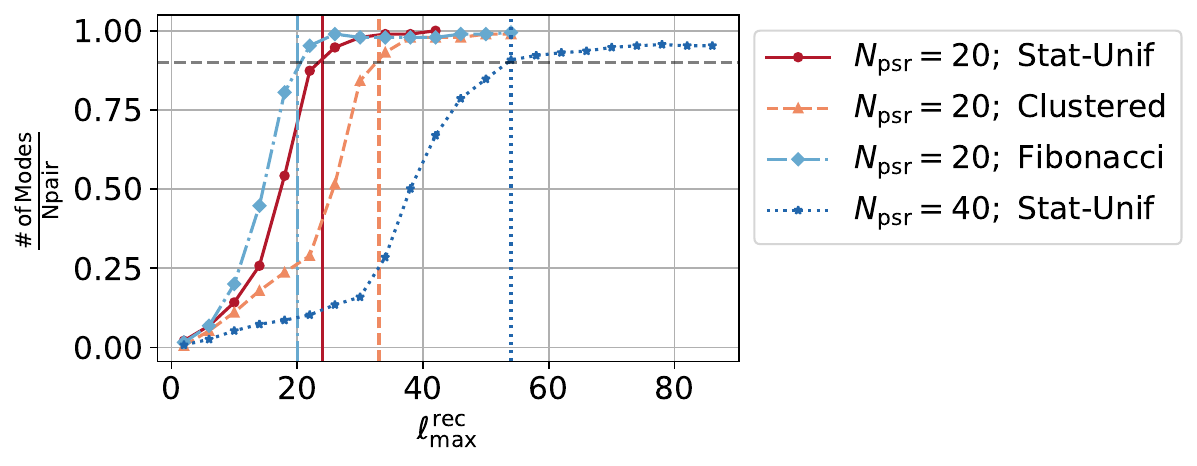}\\
    \caption{The fraction of eigenmodes with match in the range (0.95,1) normalized by $N_{\rm pair}$ as a function of $\lrec$ for the four PTA configurations shown in Fig.~\ref{f:20pulsars+}. 
    The shape of the curve depends on specifics of the pulsar configuration and flattens beyond a characteristic scale. 
    Vertical lines mark $\lres$, the maximum informative angular scale for each configuration. They have values 20, 24, 33, and 54 corresponding to the 20:Fibonacci, 20:Stat-Unif, 20:Clustered, and 40:Stat-Unif configurations, respectively.
    See also the 3rd column of Table~\ref{tab:lmax_res}.}
    \label{fig:DifferentPTA_lres}
\end{figure*}
For each configuration, we follow the same procedure as in the previous section to identify $\lres$. First, we find that the observable space is spanned by SpH basis functions up to $\lres>\lNpair$ in all cases. Second, the value of $\lres$ depends sensitively on the pulsar sky geometry and the total number of pulsars. 
These results are shown in Fig.~\ref{fig:DifferentPTA_lres} and are summarized in Tab.~\ref{tab:lmax_res}.
Figure~\ref{fig:DifferentPTA_lres} shows the fraction of eigenmodes whose match with the true eigenvectors lies in the range (0.95,1) as a function of $\lrec$.
(See right panel of Fig.~\ref{f:34eigenvector} for a similar plot for the $N_{\rm psr}=34$, Stat-Unif configuration analyzed in Sec.~\ref{s:lres}.)
\begin{table*}[htb!]
    \centering
    \begin{tabular}{c|c|c|c|c|c|c}
    \hline
    \hline
    ~~\multirow{3}{*}{PTA Configuration}~~&~~ \multirow{3}{*}{$\lNpair$}~~ &\multicolumn{2}{c|}{~~Eigenspace-based~~} &    \multicolumn{2}{c|}{Pulsar-geometry-based}  &  ~~ \multirow{3}{*}{$\dfrac{\lres}{\lnn}$}~~\\
    \cline{3-6}
         &~~ ~~ &~~ \multirow{2}{*}{$\lres$}~~ &~~ \multirow{2}{*}{$\leff$}~~ &~~ $\delta_{nn}$  ~~&~~ \multirow{2}{*}{$\lnn={\rm round}\left(\dfrac{180^\circ}{\delta_{nn}}\right)$}~~ &  \\
        &  & & & (degrees) &   \\
    \hline
       20 Stat-Unif & 12  & 24 & 10 & 15.1 & 12  & 2\\
       20 Clustered & 12  & 33 & 6 & 12.1 & 15 & 2.2\\
       20 Fibonacci & 12  & 20 & 10 & 40.4 & 4   & 5\\
      40 Stat-Unif &  26   & 54 & 20 & 6.89 & 26 & 2.07\\ 
      \hline
      \hline
    \end{tabular}
    \caption{Summary of PTA configurations and their corresponding angular resolution metrics. For each configuration, we list: (i) the counting-argument-based limit of the maximum reconstructible multipole, $\lNpair$; (ii) results from the Fisher eigenspace-based analysis---specifically, the maximum informative multipole $\lres$ and effective multipole $\leff$; and (iii) pulsar-geometry-based quantities, including the angular separation $\delta_{nn}$ (in degrees) of the nearest-neighbor pulsar pair, and its associated multipole scale $\lnn \equiv 180/\delta_{nn}$; and (iv) the ratio $\ell_{\rm max}^{\rm res}/\lnn$. Together, these quantities illustrate how pulsar geometry affects the effective angular scales accessible to a PTA.}
    \label{tab:lmax_res}
\end{table*}

\begin{figure*}[htbp!]
\centering
\includegraphics[scale=0.5]{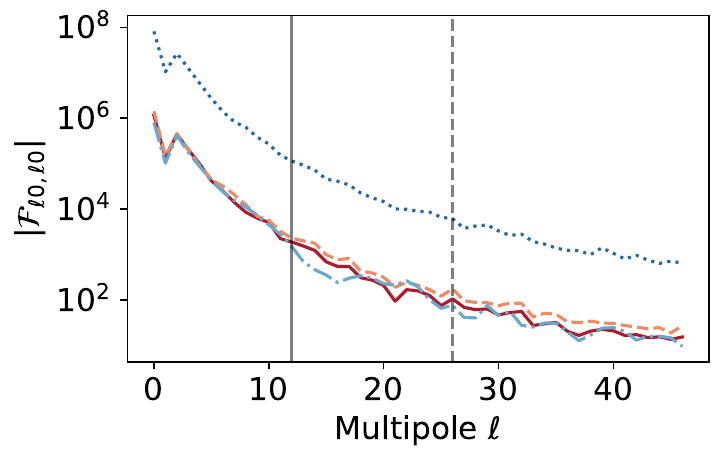}
\includegraphics[scale=0.5]{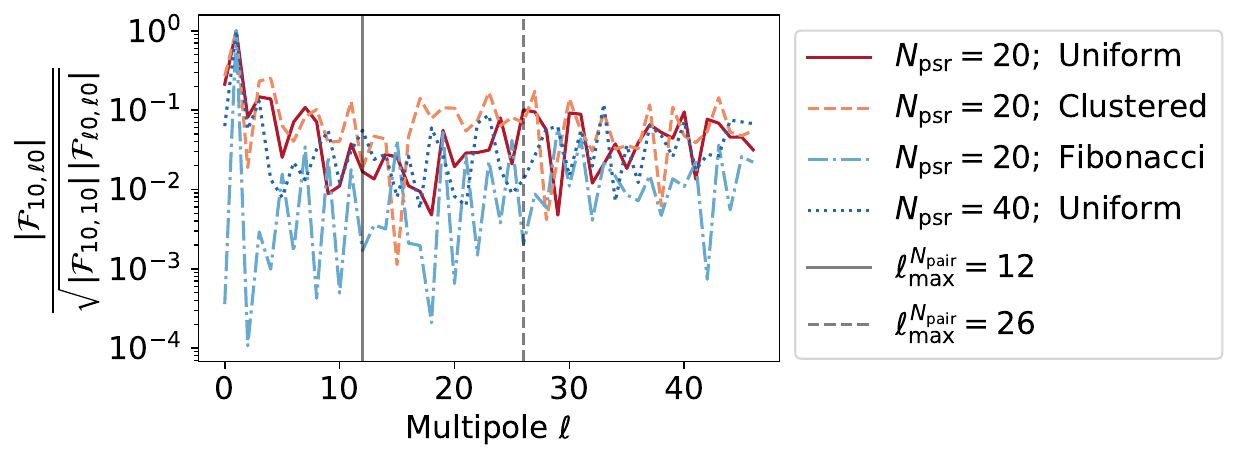}
    \caption{Plots showing mode suppression (left panel) and mode coupling (right panel) for the different PTA configurations shown in Fig.~\ref{f:20pulsars+}, similar to those shown in Fig.~\ref{f:34suppression_coupling}.}
    \label{fig:DifferentPTA_suppression}
\end{figure*}

Comparing the $N_{\rm psr}=20$, Stat-Unif and $N_{\rm psr}=20$, Clustered configurations, we find that $\lres$ is larger for the clustered case. This can be attributed to two effects: (i) the minimum pulsar separation angle for the clustered configuration is smaller than that for the statistically uniform configuration; and (ii) the concentration of pulsars in one hemisphere leads to enhanced sensitivity over a portion of sky. 
However, this increased sensitivity comes at the cost of greater nonuniformity in the pulsar array response, which in turn induces stronger multipole coupling (see right panel of Fig.~\ref{fig:DifferentPTA_suppression}) and smaller effective multipole $\leff$.

Comparing the $N_{\rm psr}=20$, Stat-Unif and $N_{\rm psr}=20$, Fibonacci configurations, we find that $\lNpair$ for the Fibonacci configuration provides a much better representation of the observable space and lies closer to $\lres$ than it does for the statistically uniform configuration. 
This is because the Fibonacci lattice yields a nearly uniform sky coverage, resulting in weaker multipole coupling (see right panel of Fig.~\ref{fig:DifferentPTA_suppression}). 
We find that the effective multipole is $\leff=10$ for both configurations considered.

Next, we compare the $N_{\rm psr}=20$, Stat-Unif and $N_{\rm psr}=40$, Stat-Unif configurations. 
As expected, the configuration with a larger number of pulsars results in higher $\lres$ and $\leff$. 
This follows from the increased overall sensitivity (left panel of Fig.~\ref{fig:DifferentPTA_suppression}) and smaller angular separations between pulsars, which allow the array to probe finer angular structures.

Thus, we see that for different pulsar configurations having the same number of pulsars (i.e., $N_{\rm psr}=20$), the Fibonacci configuration requires the smallest $\lrec$ to achieve a high fraction, whereas the clustered configuration requires the largest.

Finally, we address whether $\lres$ can be estimated using simple geometric considerations, such as the angular separation $\delta_{nn}$ between the nearest-neighbor pulsar pair, discussed also in \cite{Grunthal2026}. 
In the noise-free case---where the combined geometric response plays the dominant role---we observe an empirical relation, which is summarized in Tab.~\ref{tab:lmax_res}. 
Specifically, the ratio between $\lmax$ values corresponding to the angular separation of the nearest-neighbor pulsar pair and $\lres$ lies in the range $\sim 2$-5. This suggests that basic geometric metrics may provide useful guidance in selecting $\lrec$ for high-precision pulsar timing experiments, where pulsar geometry is expected to play a more significant role than noise. Establishing and understanding this empirical relationship on a firmer theoretical footing is left for future work.

\section{Summary / discussion}
\label{s:summary}

In this work, we identified a maximum informative angular scale $\lres$ by exploiting the information encoded in the Fisher information matrix, and we demonstrated that the commonly used counting-based limit on the maximum reconstructible multipole, $\lNpair \sim \sqrt{N_{\rm pair}}-1$, does not fully capture the true angular resolution of a PTA. 

Using both numerical simulations and analytical calculations, we confirmed 
that truncating spherical harmonic expansions at $\lrec<\lres$ leads to leakage from unmodeled yet informative modes, producing the positive bias in the recovered angular power spectrum reported in \cite{Semenzato:2025sqc}. 
We showed that choosing $\lrec\geq\lres$ eliminates this leakage-induced bias and increases the number of multipoles recovered with reduced bias. 
We also demonstrated that GWB power for angular scales beyond $\lres$ does not significantly affect the reconstruction, consistent with the PTA response acting as a low-pass filter. 

We identified a negative bias in the reconstructed angular power spectrum, which reflects the fundamental loss of information due to the finite angular resolution of the array and the resulting blurring of fine-scale sky features. 
We attempted to mitigate this effect by constructing a debiased estimator that recovers information from the partially lost modes. 
While using such an estimator statistically reduced the bias, the variance grows rapidly, signaling the presence of fundamentally unrecoverable multipoles. 

We find that due to information loss at small angular scales and mode coupling, the effective number of recoverable modes is smaller than $\lres$, which we quantify via an effective multipole $\leff$. 
The values of $\lres$ and $\leff$ depend on specifics of the pulsar sky distribution.
In Sec.~\ref{s:other-PTA-configs} we further related $\lres$ empirically to $\lnn$, constructed from the angular separation of the nearest-neighbor pulsar pair, providing a geometric estimate of a PTA's maximum informative angular scale.

In the present study, we adopted a ``hybrid'' approach: the formalism is derived in the noise-dominated limit to motivate the origin of key equations and, most importantly, the form of the Fisher information matrix. 
We then studied the behavior of $\lres$ using simulations for a simplified toy model that ignored the presence of noise, in order to establish a best-case (upper-bound) scenario and identify fundamental limiting factors imposed by pulsar geometry alone. This approach also ensures consistency with the noiseless case presented in~\cite{Semenzato:2025sqc}. Extending this analysis to realistic PTA configurations, including varying pulsar noise, remains an important direction for future work.

Finally, it would also be interesting to further investigate the observed negative bias and alternative methods for constructing unbiased estimators.
A more detailed study, involving additional simulations, will be necessary to fully characterize the behavior and performance of such estimators. 
We plan to explore these questions in forthcoming studies.

Some of the results in this paper have been derived using the {\tt healpy} and {\tt HEALPix} packages.

\begin{acknowledgments}
We thank Nicola Bellomo, Chiara Mingarelli, Federico Semenzato, Stephen R. Taylor, and Eric Thrane for friendly and fruitful discussions. D.A.~and J.D.R.~acknowledge financial support from 
NSF Physics Frontier Center Award No. PFC-2020265 and 
start-up funds from the University of Texas Rio Grande Valley.
\end{acknowledgments}

\section*{DATA AVAILABILITY}

There are no publicly available research data or software supporting this manuscript. Requests for further information or data should be sent to the authors.

%\begin{appendices}
\appendix

\section{Statistical properties of the frequentist estimators}
\label{s:stats_prop}

For completeness, we write down the expectation values and covariances between some key frequentist estimators discussed in the main text:

\begin{enumerate}

\item
Cross-correlations~\cite{Pol:2022sjn}:
\be
\ba
&\langle \rho_{ab} \rangle = \sum_{\barl =0}^{\ell_{\rm max}^{\rm gwb}}\sum_{\barm=-\barl}^{\barl}\gamma_{ab,\barl \barm}\, P_{\barl \barm}\,,\\
&\langle \rho_{ab} \rho_{cd} \rangle - \langle \rho_{ab} \rangle\langle\rho_{cd} \rangle = \delta_{ac}\, \delta_{bd}\, \sigma^2_{ab}\,.
\label{eq:app_rho_expectation}
\ea
\ee

\item 
Dirty map:
\be
\ba
&\langle X_{\ell m} \rangle =\sum_{\bar\ell \barm} \sum_{a<b} \frac{\gamma^*_{ab,\ell m}\,\gamma_{ab,\barl \barm}}{\sigma^2_{ab}}\,P_{\barl \barm}
=\sum_{\bar\ell \barm} \mathcal{F}_{\ell m,\bar\ell \barm} P_{\bar\ell \barm}
\,,\\
&\langle X_{\ell m}\,X^*_{\ell' m'} \rangle - \langle X_{\ell m}\rangle\,\langle X^*_{\ell' m'} \rangle = \mathcal{F}_{\ell m,\ell' m'}\,,
\label{eq:dirty_expectation}
\ea
\ee
which follow from \eqref{eq:rho_ab_expectation},~\eqref{eq:dirty_fisher}, and~\eqref{eq:app_rho_expectation}.

\item 
Clean map:
\be
\ba
&\langle \hat{P}_{\ell m} \rangle =\sum_{\bar\ell \barm} K_{\ell m,\bar\ell \barm} \,P_{\bar\ell \barm}\,,\\
&\langle \hat{P}_{\ell m}\,\hat{P}^*_{\ell' m'} \rangle - \langle \hat{P}_{\ell m} \rangle\,\langle \hat{P}^*_{\ell' m'} \rangle ={\mathcal{N}}_{\ell m,\ell'm'}\,,
\label{eq:clean_expectation}
\ea
\ee
where
\begin{align}
\label{e:K}
&K_{\ell m,\bar\ell \barm} \equiv \sum_{\ell' m'} \mathcal{F}^{+}_{\ell m, \ell' m'}\, \mathcal{F}_{\ell' m',\barl\barm}\,,
\\
&
\label{e:calN}
{\mathcal{N}}_{\ell m,\ell'm'}\equiv(\Fmat^+\cdot\Fmat\cdot\Fmat^{+\dagger})_{\ell m,\ell'm'}\,,
\end{align}
which follow from
\eqref{eq:pseudo_clean_estimator}
and~\eqref{eq:dirty_expectation}.
Note that $\bm{K}$ is a (nonsquare) rectangular matrix if $\ell_{\rm max}^{\rm rec}\neq \ell_{\rm max}^{\rm gwb}$.

\item
Noise bias of standard angular power spectrum estimator $\hat C_{\ell}$:
\be 
\ba
\llangle \hat C_{\ell} \rrangle_P&\equiv\frac{1}{2\ell+1}\, \sum_{m} \llangle \hat{P}_{\ell m}\,\hat{P}^*_{\ell m}\rrangle_P-N_{\ell}\\
& =\frac{1}{2\ell+1}\sum_{m}\left[\llangle \hat{P}_{\ell m} \rangle\langle \hat{P}^*_{\ell m} \rrangle_P +{\mathcal{N}}_{\ell m,\ell m}\right]-N_{\ell}
\\
&=\frac{1}{2\ell+1}\sum_{m}\,\llangle \hat{P}_{\ell m} \rangle\langle \hat{P}^*_{\ell m} \rrangle_P   \\
& =\frac{1}{2\ell+1}\sum_{m}\, \sum_{\barl \barm}\sum_{\barl' \barm'} K_{\ell m,\barl\barm}\, K^*_{\ell m,\barl'\barm'} \,\langle P_{\barl \barm}  \,P^*_{\barl' \barm'} \rangle_P 
\\
& =\frac{1}{2\ell+1}\sum_{m}\, \sum_{\barl \barm}\sum_{\barl' \barm'} K_{\ell m,\barl\barm}\, K^*_{\ell m,\barl'\barm'} \,C_{\barl}\,\delta_{\barl\barl'}\,\delta_{\barm \barm'}
\\
&=\sum_{\barl} M_{{\ell} \barl} \,C_{\barl}\,,
\label{eq:noise_Bias}
\ea
\ee
where 
\be
N_{\ell}
\equiv \frac{1}{2\ell + 1}
\sum_{ m}
{\mathcal{N}}_{\ell m,\ell m}\,,\qquad
M_{\ell \barl} \equiv\frac{1}{2\ell+1}\sum_{m \barm} |K_{\ell m,\barl\barm}|^2\,.
\label{eq:bias_matrix}
\ee
Similar to $\bm{K}$, the bias matrix $\bm{M}$ is a (nonsquare) rectangular matrix if $\ell_{\rm max}^{\rm rec}\neq \ell_{\rm max}^{\rm gwb}$.

\i Covariance of standard angular power spectrum estimator $\hat C_{\ell}$:
\be
\label{eq:Cl_cov}
{\rm Cov}(\hat{C}_{\ell}, \hat{C}_{\ell'})
=\frac{2}{(2\ell+1)(2\ell'+1)}\sum_{mm'} |{\cal C}_{\ell m,\ell' m'}+{\cal N}_{\ell m,\ell' m'}|^2\,,
\ee
where
\be
{\cal C}_{\ell m,\ell' m'}\equiv\sum_{\barl \barm} K_{\ell m,\barl \barm }\,K^*_{\ell' m',\barl \barm}\,C_{\barl}\,,
\label{e:Clml'm'}
\ee
and ${\cal N}_{\ell m,\ell' m'}$ is given in \eqref{e:calN}.

An outline of a proof of \eqref{eq:Cl_cov} is as follows:
\ben

\i It is convenient to first write
\be
\hat P_{\ell m}
\equiv \mu_{\ell m}+n_{\ell m}\,,
\quad{\rm where}\quad
\mu_{\ell m}
\equiv \langle \hat P_{\ell m}\rangle =\sum_{\bar\ell \barm} K_{\ell m,\bar\ell \barm} \,P_{\bar\ell \barm}
\ee
and $n_{\ell m}$ is a Gaussian random variable satisfying
\be
\langle n_{\ell m}\rangle=0\,,
\qquad
\langle n_{\ell m}n^*_{\ell' m'}\rangle={\cal N}_{\ell m,\ell' m'}\,.
\label{e:nstats}
\ee

\i Using \eqref{eq:stats_Iso}, one can show that
\be
\langle \mu_{\ell m}\rangle_P=0\,,
\qquad
\langle \mu_{\ell m}\mu^*_{\ell' m'}\rangle_P=\sum_{\barl \barm} K_{\ell m,\barl \barm }\,K^*_{\ell' m',\barl \barm}\,C_{\barl}={\cal C}_{\ell m,\ell' m'}\,.
\label{e:mustats}
\ee

\i Note that $\mu_{\ell m}$ is a deterministic quantity relative to the $\langle\ \rangle$  expectation values implying that $\mu_{\ell m}$ and $n_{\ell m}$ are statistically independent of one another. (Note also that $n_{\ell m}$ is a deterministic quantity with respect to the $\langle\ \rangle_P$ expectation values, but this property is not needed below to calculate the covariance.) 

\i Then given the estimator 
\be
\hat C_{\ell} = \frac{1}{2\ell l+1}\sum_{m}|\hat P_{\ell m}|^2 - N_{\ell}\,,
\ee
we can calculate the covariance
\be
{\rm Cov}(\hat{C}_{\ell}, \hat{C}_{\ell'})
\equiv\llangle\hat C_{\ell}\,\hat C_{\ell'}\rrangle_P-
\llangle\hat C_{\ell}\rrangle_P
\llangle\hat C_{\ell'}\rrangle_P
\ee
by first expanding the rhs in terms of expectation values of 4th-order and 2nd-order products of $\hat P_{\ell m}$ and then further expanding those expectation values in terms products of $\mu_{\ell m}$ and $n_{\ell m}$.

\i Expectation values $\langle\ \rangle$ of products of 
$\mu_{\ell m}$ and $n_{\ell m}$ involving 
an odd number of $\mu$'s or $n$'s vanish.
Expectation values of quadratic or 4th-order products of $\mu$'s or $n$'s can be evaluated using \eqref{e:mustats} and \eqref{e:nstats} and Isserlis's theorem~\cite{isserlis:1918} (to reduce 4th-order expectation values to products of 2nd-order expectation values).
This leads to terms involving products of ${\cal N}_{\ell m,\ell' m'}$ and ${\cal C}_{\ell m,\ell' m'}$.

\i Finally, these terms can be combined using the Hermitian properties of ${\cal N}_{\ell m,\ell' m'}$ and ${\cal C}_{\ell m,\ell' m'}$:
\be
{\cal N}_{\ell m,\ell' m'}
={\cal N}^*_{\ell' m',\ell m}
=
(-1)^{m+m'}
{\cal N}^*_{\ell,-m;\ell',-m'}
\ee
and similarly for ${\cal C}_{\ell m,\ell' m'}$,
leading to the final result given in \eqref{eq:Cl_cov}.

\een

\i Debiased angular power spectrum estimator:
\be
\label{e:debiased_estimator}
\hat{C}_{\ell}^{(u)} \equiv \sum_{\ell'}\,M^+_{\ell\ell'}\,\hat{C}_{\ell'} \,,
\ee
has expected value and covariance matrix
\begin{align}
\label{e:debiased_mean_app}
&\llangle \hat C^{(u)}_{\ell}\rrangle_P
=\sum_{\ell'}\sum_{\barl}
M^+_{\ell\ell'}
M^{\phantom{+}}_{\ell'\barl}C_{\barl}\,,
\\
\label{eq:debiased_cov_app}
&{\rm Cov}(\hat C^{(u)}_{\ell}, \hat C^{(u)}_{\ell'})=\sum_{\ell_1}\sum_{\ell_2}\frac{2}{(2\ell_1+1)(2\ell_2+1)}\sum_{m m'} M^+_{\ell \ell_1}|{\cal C}_{\ell_1 m,\ell_2 m'}+{\cal N}_{\ell_1 m,\ell_2 m'}|^2 M^+_{\ell_2 \ell'}\,,
\end{align}
where the last two equalities follow from \eqref{eq:noise_Bias} and \eqref{eq:Cl_cov}.
\end{enumerate}

\section{Impact of unmodeled angular scales in the absence of mode coupling}
\label{app:noBiasNoCoupling}

In the main text, we claimed that it is the {\it combination} of both mode coupling and unmodeled informative angular scales (i.e., choosing $\lrec<\lres$) that leads to leakage. 
To justify that claim, we explicitly show here that if there were no mode coupling, then there would be no leakage even if we choose $\lrec<\lres$.

So, let us reconsider the case of a dense-PTA Fisher matrix as in~\eqref{e:gaussian_beam_harmonic}.
For this case, there is no mode coupling, meaning that the Fisher matrix is diagonal in harmonic space:
\be
\mathcal{F}_{\ell m,\ell' m'}=\mathcal{F}_{\ell}\,\delta_{\ell\ell'}\,\delta_{mm'}\,.
\label{e:diagF}
\ee
Here the indices $\ell$ and $m$ range over there full range of values for the PTA, i.e., $0\le\ell\le\lres$.
We will assume that $\mathcal{F}_\ell\ne 0$ for all $\ell$, so that the pseudoinverse of the Fisher matrix is just the ordinary inverse:
\be
(\mathcal{F}^+)_{\ell m,\ell' m'}=
(\mathcal{F}^{-1})_{\ell m,\ell' m'}=\mathcal{F}_{\ell}^{-1}\,\delta_{\ell\ell'}\,\delta_{mm'}\,,
\label{e:F+=F-1}
\ee
From \eqref{e:K}, it follows that
\be
\ba
K_{\ell m, \barl\barm}
&\equiv  \sum_{\ell' m'} \mathcal{F}^{+}_{\ell m, \ell' m'}\,\mathcal{F}_{\ell' m',\barl\barm}
\\
&=  \sum_{\ell' m'} \mathcal{F}^{-1}_{\ell m, \ell' m'}\,\mathcal{F}_{\ell' m',\barl\barm}
\\
&= \sum_{\ell' m'} \mathcal{F}^{-1}_{\ell}
\delta_{\ell\ell'}\delta_{mm'}\,
\mathcal{F}_{\ell'} \delta_{\ell'\barl} \delta_{m'\barm}
\\
&= 
\delta_{\ell\barl}\delta_{m\barm}
\ea
\ee
where the second equality follows from the first equality in \eqref{e:F+=F-1};
the third equality follows from \eqref{e:diagF} and
\eqref{e:F+=F-1}; 
the fourth equality comes from using the Kronecker deltas 
$\delta_{\ell\ell'}\delta_{mm'}$ to eliminate the sum over $\ell'm'$.
Then the bias matrix $\bm{M}$ defined in \eqref{eq:bias_matrix} becomes
\be
M_{\ell\barl}
\equiv\frac{1}{2\ell+1} \sum_{m \barm} |K_{\ell\barl,m \barm}|^2
=\frac{1}{2\ell+1} \sum_{m \barm} \big|\delta_{\ell\barl}\delta_{m \barm}\big|^2
=\frac{1}{2\ell+1} \sum_{m}\delta_{\ell\barl}
=\delta_{\ell\barl}\,.
\ee
Finally, by inspection of \eqref{eq:noise_Bias} and \eqref{eq:bias_Cl_estimator_1}, we see that $b_{\ell}=0$.
Thus, there is no bias if there is no coupling, even if we choose $\lrec<\lres$.

\section{Impact of choice of condition number on $C_\ell$ reconstruction}
\label{s:condition_number}

As mentioned in the main text, both the standard and debiased estimators ($\hat C_\ell$ and $\hat C^{(u)}_\ell$) of the angular power spectrum involve pseudoinverses in their definitions.
Since calculating a pseudoinverse depends on the choice of a condition number threshold, so too do the corresponding estimators.
In this Appendix, we demonstrate the impact of the choice of condition number on the recovery of the $C_\ell$'s.

\subsection{Standard estimator}
\label{s:impact_standard}

We start by discussing the standard estimator $\hat C_\ell$, defined by \eqref{eq:Cl_estimator_1}, which requires inverting the Fisher information matrix \eqref{eq:dirty_fisher}.
Since the Fisher matrix is not invertible, we use its pseudoinverse $\mb{\Fmat}^+$ given by \eqref{e:SVD_2} and \eqref{e:SVD_3}.
The bias and variance of the standard estimator $\hat C_\ell$ can be calculated using \eqref{eq:bias_Cl_estimator_1} and \eqref{eq:Cl_cov_main}.

\begin{figure*}[htb!]
\centering
\includegraphics[width=0.49\textwidth]
{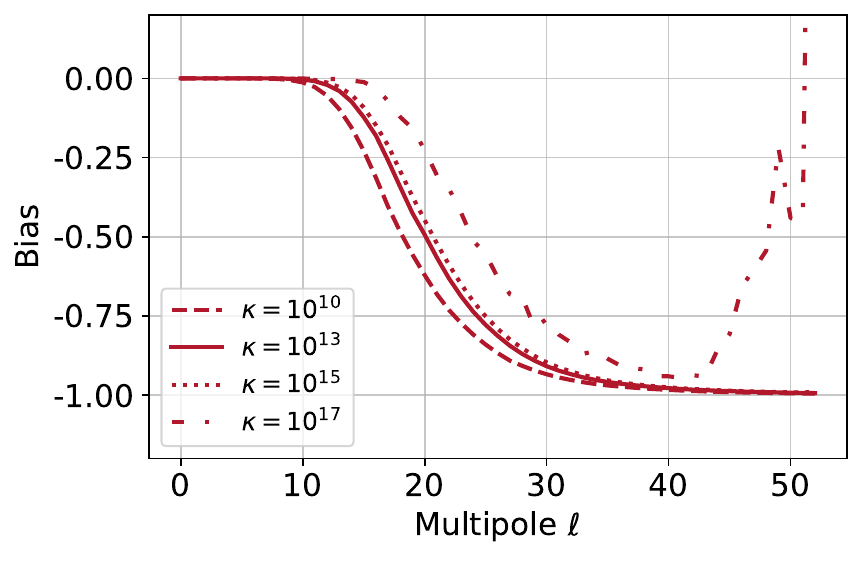}
\includegraphics[width=0.49\textwidth]
{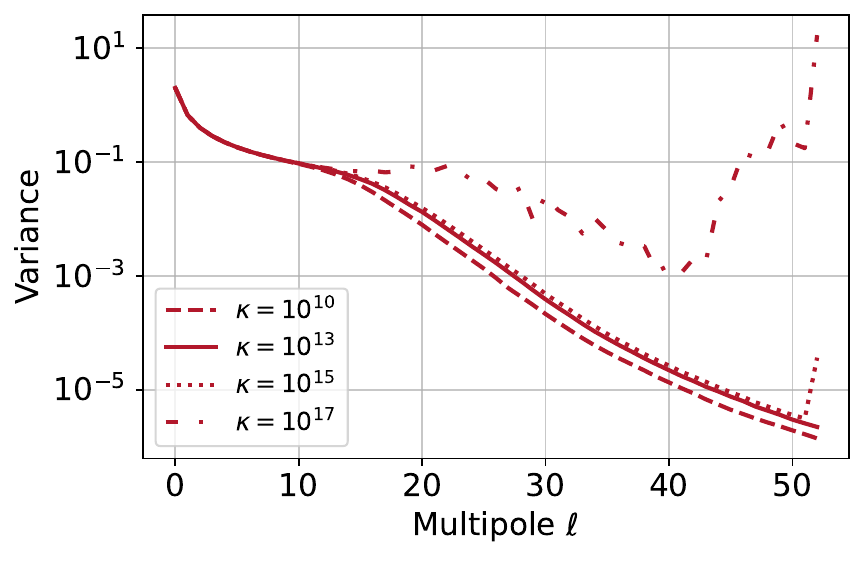}
\caption{
Dependence of the bias (left panel) and variance (right panel) of the standard estimator $\hat C_\ell$ on the choice of condition number used for calculating the pseudoinverse of the Fisher matrix.
For these plots, we consider the case $\ell_{\rm max}^{\rm rec}=52$ and $\ell_{\rm max}^{\rm gwb}=52$.
The negative bias (left panel) reflects information loss due to the finite ``width'' of the Fisher matrix (or point spread function) in position space. 
For $\kappa=10^{17}$, we observe somewhat erratic changes in the bias and variance, which we attribute to the amplification of numerical errors.
The analyses presented in Fig.~\ref{f:34Clrecovery_top} correspond to the choice $\kappa=10^{13}$.}
\label{f:standard_analyticVar}
\end{figure*}

The left and right panels of Fig.~\ref{f:standard_analyticVar} illustrate the dependence of the bias and variance, respectively, of $\hat C_\ell$ on the choice of condition-number threshold. 
The plots in this figure correspond to the case where $\ell_{\rm max}^{\rm rec}=52$ and $\ell_{\rm max}^{\rm gwb}=52$.
As the condition-number threshold increases from $\kappa=10^{10}$ to $\kappa=10^{15}$, the bias decreases while the variance increases. 
For $\kappa=10^{17}$, we observe increases in both bias and variance, which we attribute to the amplification of numerical errors.
For reference, the analyses presented in Fig.~\ref{f:34Clrecovery_top} correspond to the choice $\kappa=10^{13}$.

\subsection{Debiased estimator}
\label{s:impact_debiased}

We next discuss the impact of the choice of condition number on $C_\ell$ reconstruction when using the debiased standard estimator $\hat C^{(u)}_\ell$ defined in \eqref{e:debiased_estimator}.
For this discussion we will restrict our attention to $\lrec=52$, which requires choosing a condition number for two matrix inversions.
That is, in addition to choosing a condition number to invert the Fisher matrix $\bm{\Fmat}$ (as described in the previous subsection), we must also choose a condition number to invert the bias matrix $\bm{M}$, which is shown in the left panel of Fig.~\ref{f:34Mmatrix}.
Here, we will fix $\kappa=10^{13}$ for inverting the Fisher matrix (as described above), and 
examine four different choices of the condition-number threshold for inverting $\bm{M}$, which are $\{10,10^2,10^3,10^4\}$.

\begin{figure*}[htb!]
\centering
\includegraphics[width=\textwidth]{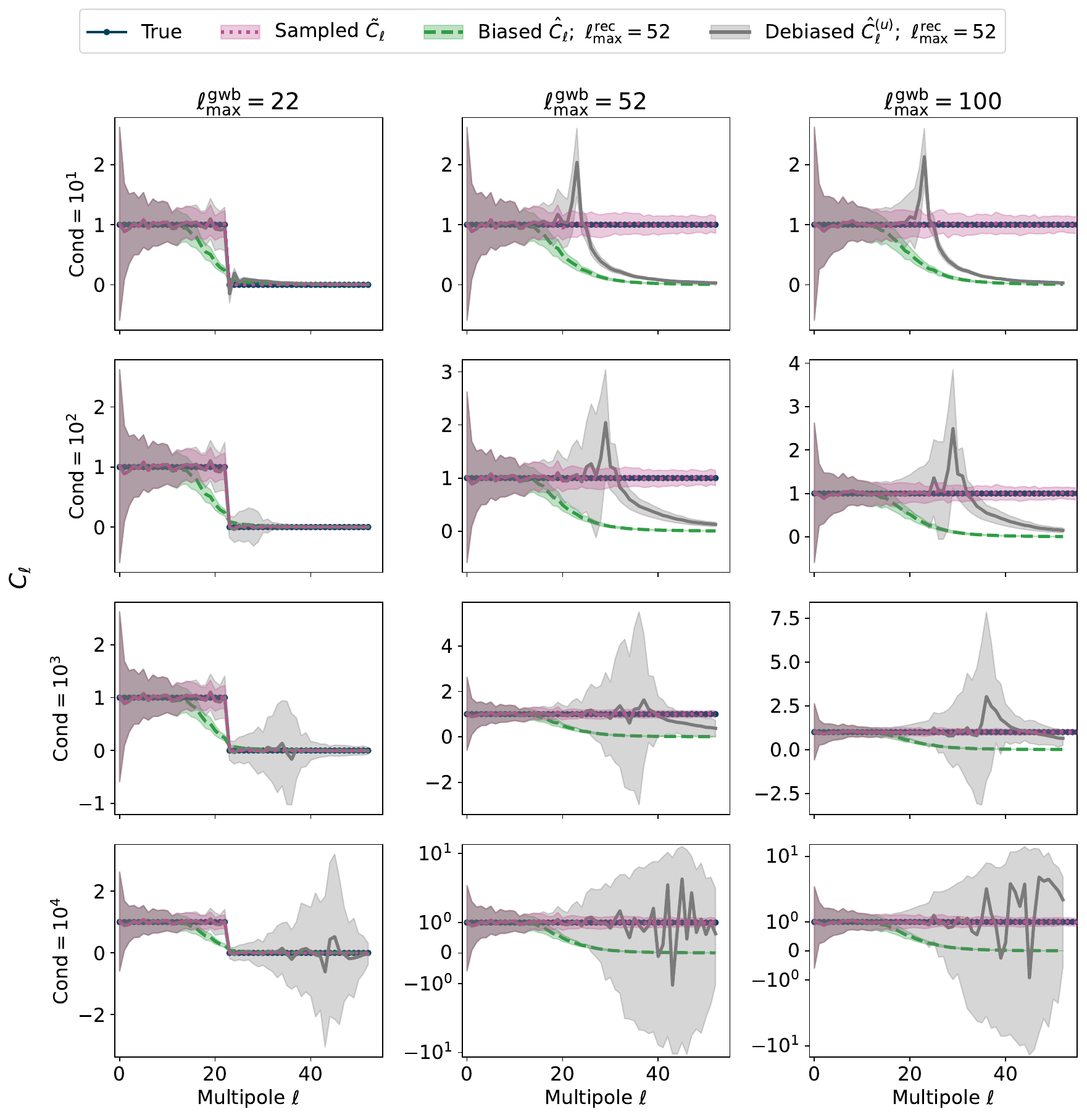}
    \caption{Similar to Fig.~\ref{fig:debiased_52} (for $\lrec=52$) for different values of the condition-number threshold $\{10,10^2,10^3,10^4\}$ (different rows, top to bottom) used to construct the pseudoinverse for the bias matrix $\bm{M}$.
    The mean recovered $C_\ell$ with $\pm 1\sigma$ uncertainties (shaded bands) are compared to the injected spectrum (blue solid circle). 
    Debiasing improves agreement with the injected spectrum for well-resolved multipoles ($\lesssim 22$). 
    As the threshold increases, the bias reduces but the variance increases for poorly constrained modes ($>22$).}
\label{fig:debiased_52_vs_condition}
\end{figure*}

The results are presented in Fig.~\ref{fig:debiased_52_vs_condition}, where rows from top to bottom correspond to increasing condition-number threshold for $\bm{M}$. 
We also consider three different values of $\lgwb\in\{22, 52, 100\}$, corresponding to the three different columns in the plot.
For a condition-number threshold of $10$ (top row), the debiased estimator shows clear improvement relative to the standard estimator, with approximately five additional multipoles consistent with the injected spectrum within $\pm 1\sigma$ uncertainty (top middle panel). 
However, as the condition number threshold is increased (lower rows), we find that although the debiased estimator formally starts to become statistically unbiased, the variance grows rapidly, leading to increasingly noisy reconstructions. 
This behavior indicates that modes beyond $\ell\approx 22$ are intrinsically difficult to recover.

\bibliography{ref}

@article{Hellings-Downs:1983,
        adsurl = {https://ui.adsabs.harvard.edu/abs/1983ApJ...265L..39H},
        author = {{Hellings}, R.~W. and {Downs}, G.~S.},
        doi = {10.1086/183954},
        journal = {Astrophys.~J.},
        month = feb,
        pages = {L39-L42},
        title = {{Upper limits on the isotropic gravitational radiation background from pulsar timing analysis}},
        volume = {265},
        year = 1983}

@article{isserlis:1918,
        author = {Isserlis, L.},
        doi = {10.1093/biomet/12.1-2.134},
        issn = {0006-3444},
        journal = {Biometrika},
        month = {11},
        number = {1-2},
        pages = {134-139},
        title = {{On a formula for the product-moment coefficient of any order of a normal frequency distribution in any number of variables}},
        url = {https://doi.org/10.1093/biomet/12.1-2.134},
        volume = {12},
        year = {1918}}

@article{Ali-Haimoud:2020ozu,
    title = {Fisher formalism for anisotropic gravitational-wave background searches with pulsar timing arrays},
  author = {Ali-Ha{\"i}moud, Yacine and Smith, Tristan L. and Mingarelli, Chiara M. F.},
  journal = {Phys. Rev. D},
  volume = {102},
  issue = {12},
  pages = {122005},
  numpages = {18},
  year = {2020},
  month = {Dec},
  publisher = {American Physical Society},
  doi = {10.1103/PhysRevD.102.122005},
  url = {https://link.aps.org/doi/10.1103/PhysRevD.102.122005}
}

@article{alihaimoud2021,
  title = {Insights into searches for anisotropies in the nanohertz gravitational-wave background},
  author = {Ali-Ha{\"i}moud, Yacine and Smith, Tristan L. and Mingarelli, Chiara M. F.},
  journal = {Phys. Rev. D},
  volume = {103},
  issue = {4},
  pages = {042009},
  numpages = {22},
  year = {2021},
  month = {Feb},
  publisher = {American Physical Society},
  doi = {10.1103/PhysRevD.103.042009},
  url = {https://link.aps.org/doi/10.1103/PhysRevD.103.042009}
}

@article{Semenzato:2025sqc,
    author = "Semenzato, Federico and Bellomo, Nicola and Raccanelli, Alvise and Mingarelli, Chiara M. F.",
    title = "{Bias from small-scale leakage in Pulsar Timing Array maps}",
    eprint = "2510.24857",
    archivePrefix = "arXiv",
    primaryClass = "astro-ph.IM",
    month = "10",
    year = "2025"
}

@article{Pol:2022sjn,
    author = "Pol, Nihan and Taylor, Stephen R. and Romano, Joseph D.",
    title = "{Forecasting Pulsar Timing Array Sensitivity to Anisotropy in the Stochastic Gravitational Wave Background}",
    eprint = "2206.09936",
    archivePrefix = "arXiv",
    primaryClass = "astro-ph.HE",
    doi = "10.3847/1538-4357/ac9836",
    journal = "Astrophys. J.",
    volume = "940",
    number = "2",
    pages = "173",
    year = "2022"
}

@article{Grunthal2026,
  author  = {Grunthal, Kathrin and Champion, David J. and Thrane, Eric and Nathan, Rowina S. and Kramer, Michael and Miles, Matthew T.},
  title   = {Optimising gravitational-wave sky maps for pulsar timing arrays},
  journal = {Astronomy \& Astrophysics},
  year    = {2026},
  note    = {In press},
  doi     = {10.1051/0004-6361/202557896}
}

@article{Zonca2019,
  doi = {10.21105/joss.01298},
  url = {https://doi.org/10.21105/joss.01298},
  year = {2019},
  month = mar,
  publisher = {The Open Journal},
  volume = {4},
  number = {35},
  pages = {1298},
  author = {Andrea Zonca and Leo Singer and Daniel Lenz and Martin Reinecke and Cyrille Rosset and Eric Hivon and Krzysztof Gorski},
  title = {healpy: equal area pixelization and spherical harmonics transforms for data on the sphere in Python},
  journal = {Journal of Open Source Software}
}

@ARTICLE{2005ApJ...622..759G,
   author = {{G{\'o}rski}, K.~M. and {Hivon}, E. and {Banday}, A.~J. and 
	{Wandelt}, B.~D. and {Hansen}, F.~K. and {Reinecke}, M. and 
	{Bartelmann}, M.},
    title = "{HEALPix: A Framework for High-Resolution Discretization and Fast Analysis of Data Distributed on the Sphere}",
  journal = {\apj},
   eprint = {arXiv:astro-ph/0409513},
     year = 2005,
    month = apr,
   volume = 622,
    pages = {759-771},
      doi = {10.1086/427976},
   adsurl = {http://adsabs.harvard.edu/abs/2005ApJ...622..759G}
}

@article{Domcke:2025esw,
    author = "Domcke, Valerie and Franciolini, Gabriele and Pieroni, Mauro",
    title = "{Cosmic Variance in Anisotropy Searches at Pulsar Timing Arrays}",
    eprint = "2508.21131",
    archivePrefix = "arXiv",
    primaryClass = "astro-ph.CO",
    reportNumber = "CERN-TH-2025-159",
    doi = "10.1103/5jmr-4dj7",
    journal = "Phys. Rev. Lett.",
    volume = "137",
    number = "6",
    pages = "061003",
    year = "2026"
}

@article{NANOGrav:2023tcn,
    author = "Agazie, Gabriella and others",
    collaboration = "NANOGrav",
    title = "{The NANOGrav 15 yr Data Set: Search for Anisotropy in the Gravitational-wave Background}",
    eprint = "2306.16221",
    archivePrefix = "arXiv",
    primaryClass = "astro-ph.HE",
    doi = "10.3847/2041-8213/acf4fd",
    journal = "Astrophys. J. Lett.",
    volume = "956",
    number = "1",
    pages = "L3",
    year = "2023"
}

@article{Romano:2016dpx,
    author = "Romano, Joseph D. and Cornish, Neil J.",
    title = "{Detection methods for stochastic gravitational-wave backgrounds: a unified treatment}",
    eprint = "1608.06889",
    archivePrefix = "arXiv",
    primaryClass = "gr-qc",
    doi = "10.1007/s41114-017-0004-1",
    journal = "Living Rev. Rel.",
    volume = "20",
    number = "1",
    pages = "2",
    year = "2017"
}

@article{Thrane:2009fp,
    author = "Thrane, Eric and Ballmer, Stefan and Romano, Joseph D. and Mitra, Sanjit and Talukder, Dipongkar and Bose, Sukanta and Mandic, Vuk",
    title = "{Probing the anisotropies of a stochastic gravitational-wave background using a network of ground-based laser interferometers}",
    eprint = "0910.0858",
    archivePrefix = "arXiv",
    primaryClass = "astro-ph.IM",
    doi = "10.1103/PhysRevD.80.122002",
    journal = "Phys. Rev. D",
    volume = "80",
    pages = "122002",
    year = "2009"
}

@article{Agarwal23,
  title = {Angular power spectra of anisotropic stochastic gravitational wave background: Developing statistical methods and analyzing data from ground-based detectors},
  author = {Agarwal, Deepali and Suresh, Jishnu and Mitra, Sanjit and Ain, Anirban},
  journal = {Phys. Rev. D},
  volume = {108},
  issue = {2},
  pages = {023011},
  numpages = {15},
  year = {2023},
  month = {Jul},
  publisher = {American Physical Society},
  doi = {10.1103/PhysRevD.108.023011},
  url = {https://link.aps.org/doi/10.1103/PhysRevD.108.023011}
}

@article{Mitra:2007mc,
    author = "Mitra, Sanjit and Dhurandhar, Sanjeev and Souradeep, Tarun and Lazzarini, Albert and Mandic, Vuk and Bose, Sukanta and Ballmer, Stefan",
    title = "{Gravitational wave radiometry: Mapping a stochastic gravitational wave background}",
    eprint = "0708.2728",
    archivePrefix = "arXiv",
    primaryClass = "gr-qc",
    reportNumber = "IUCAA-28-07, LIGO-P070033-Z",
    doi = "10.1103/PhysRevD.77.042002",
    journal = "Phys. Rev. D",
    volume = "77",
    pages = "042002",
    year = "2008"
}

@article{EPTA_InPTA_23,
	author = {{EPTA Collaboration and InPTA Collaboration} and {others}},
	title = {The second data release from the {European Pulsar Timing Array - III}. Search for gravitational wave signals},
	DOI= "10.1051/0004-6361/202346844",
	url= "https://doi.org/10.1051/0004-6361/202346844",
	journal = {A\&A},
	year = 2023,
	volume = 678,
	pages = "A50",
}

@article{Agazie_2023,
doi = {10.3847/2041-8213/acdac6},
url = {https://dx.doi.org/10.3847/2041-8213/acdac6},
year = {2023},
month = {jun},
publisher = {The American Astronomical Society},
volume = {951},
number = {1},
pages = {L8},
author = {Gabriella Agazie and others},
title = {The {NANOGrav} 15 yr Data Set: Evidence for a Gravitational-wave Background},
journal = {The Astrophysical Journal Letters}
}

@article{Reardon_2023,
doi = {10.3847/2041-8213/acdd02},
url = {https://dx.doi.org/10.3847/2041-8213/acdd02},
year = {2023},
month = {jun},
publisher = {The American Astronomical Society},
volume = {951},
number = {1},
pages = {L6},
author = {Daniel J. Reardon and others},
title = {Search for an Isotropic Gravitational-wave Background with the {Parkes Pulsar Timing Array}},
journal = {The Astrophysical Journal Letters}}

@article{Xu_2023,
doi = {10.1088/1674-4527/acdfa5},
url = {https://dx.doi.org/10.1088/1674-4527/acdfa5},
year = {2023},
month = {jun},
publisher = {National Astromonical Observatories, CAS and IOP Publishing},
volume = {23},
number = {7},
pages = {075024},
author = {Heng Xu and others},
title = {Searching for the Nano-Hertz Stochastic Gravitational Wave Background with the {Chinese Pulsar Timing Array Data Release I}},
journal = {Research in Astronomy and Astrophysics}
}

@article{10.1093/mnras/stae2571,
    author = {Miles, Matthew T and others},
    title = {The MeerKAT Pulsar Timing Array: the first search for gravitational waves with the MeerKAT radio telescope},
    journal = {Monthly Notices of the Royal Astronomical Society},
    volume = {536},
    number = {2},
    pages = {1489-1500},
    year = {2025},
    month = {01},
    doi = {10.1093/mnras/stae2571},
    url = {https://doi.org/10.1093/mnras/stae2571},
    eprint = {https://academic.oup.com/mnras/article-pdf/536/2/1489/61215196/stae2571.pdf},
}

@article{NANOGrav:2023hvm,
    author = "Afzal, Adeela and others",
    collaboration = "NANOGrav",
    title = "{The NANOGrav 15 yr Data Set: Search for Signals from New Physics}",
    eprint = "2306.16219",
    archivePrefix = "arXiv",
    primaryClass = "astro-ph.HE",
    reportNumber = "FERMILAB-PUB-23-589-T",
    doi = "10.3847/2041-8213/acdc91",
    journal = "Astrophys. J. Lett.",
    volume = "951",
    number = "1",
    pages = "L11",
    year = "2023",
    note = "[Erratum: Astrophys.J.Lett. 971, L27 (2024), Erratum: Astrophys.J. 971, L27 (2024)]"
}

@article{Sah:2024oyg,
    author = "Sah, Mohit Raj and Mukherjee, Suvodip and Saeedzadeh, Vida and Babul, Arif and Tremmel, Michael and Quinn, Thomas R.",
    title = "{Imprints of supermassive black hole evolution on the spectral and spatial anisotropy of nano-hertz stochastic gravitational-wave background}",
    eprint = "2404.14508",
    archivePrefix = "arXiv",
    primaryClass = "astro-ph.CO",
    doi = "10.1093/mnras/stae1930",
    journal = "Mon. Not. Roy. Astron. Soc.",
    volume = "533",
    number = "2",
    pages = "1568--1582",
    year = "2024"
}

@article{Sah:2024etc,
    author = "Sah, Mohit Raj and Mukherjee, Suvodip",
    title = "{Discovering the Cosmic Evolution of Supermassive Black Holes using Nano-Hertz Gravitational Waves and Galaxy Surveys}",
    eprint = "2407.11669",
    archivePrefix = "arXiv",
    primaryClass = "astro-ph.CO",
    month = "7",
    year = "2024"
}

@article{Lin:2026owz,
    author = "Lin, Meng-Xiang and Lidz, Adam and Ma, Chung-Pei",
    title = "{Analytical estimates of gravitational wave background anisotropies from shot noise and large-scale structure in pulsar timing arrays}",
    eprint = "2602.16808",
    archivePrefix = "arXiv",
    primaryClass = "astro-ph.CO",
    doi = "10.1103/p534-cvj8",
    journal = "Phys. Rev. D",
    volume = "113",
    number = "12",
    pages = "123052",
    year = "2026"
}

@article{Semenzato:2024mtn,
    author = "Semenzato, Federico and Casey-Clyde, J. Andrew and Mingarelli, Chiara M. F. and Raccanelli, Alvise and Bellomo, Nicola and Bartolo, Nicola and Bertacca, Daniele",
    title = "{Cross-correlating the Universe: The Gravitational Wave Background and Large-Scale Structure}",
    eprint = "2411.00532",
    archivePrefix = "arXiv",
    primaryClass = "astro-ph.CO",
    doi = "10.1088/1475-7516/2026/06/066",
    journal = "JCAP",
    volume = "06",
    pages = "066",
    year = "2026"
}

@article{Grunthal:2024sor,
    author = "Grunthal, Kathrin and others",
    title = "{The MeerKAT Pulsar Timing Array: Maps of the gravitational wave sky with the 4.5-yr data release}",
    eprint = "2412.01214",
    archivePrefix = "arXiv",
    primaryClass = "astro-ph.HE",
    doi = "10.1093/mnras/stae2573",
    journal = "Mon. Not. Roy. Astron. Soc.",
    volume = "536",
    number = "2",
    pages = "1501--1517",
    year = "2024"
}

@article{Boyle:2010rt,
    author = "Boyle, Latham and Pen, Ue-Li",
    title = "{Pulsar timing arrays as imaging gravitational wave telescopes: angular resolution and source (de)confusion}",
    eprint = "1010.4337",
    archivePrefix = "arXiv",
    primaryClass = "astro-ph.HE",
    doi = "10.1103/PhysRevD.86.124028",
    journal = "Phys. Rev. D",
    volume = "86",
    pages = "124028",
    year = "2012"
}

@article{Taylor:2015udp,
    author = "Taylor, S. R. and others",
    title = "{Limits on anisotropy in the nanohertz stochastic gravitational-wave background}",
    eprint = "1506.08817",
    archivePrefix = "arXiv",
    primaryClass = "astro-ph.HE",
    doi = "10.1103/PhysRevLett.115.041101",
    journal = "Phys. Rev. Lett.",
    volume = "115",
    number = "4",
    pages = "041101",
    year = "2015"
}

@article{Chen:2026mid,
    author = "Chen, Yiqin and others",
    title = "{Searching for anisotropy in the gravitational wave background using the Parkes Pulsar Timing Array}",
    eprint = "2602.11529",
    archivePrefix = "arXiv",
    primaryClass = "astro-ph.HE",
    doi = "10.1103/czxp-zrd6",
    journal = "Phys. Rev. D",
    volume = "113",
    number = "4",
    pages = "043042",
    year = "2026"
}

@article{Lin:2026qvc,
    author = "Lin, Meng-Xiang and Lidz, Adam and Ma, Chung-Pei",
    title = "{Realization Variance of Gravitational Wave Background Anisotropies from Shot Noise for Pulsar Timing Arrays}",
    eprint = "2608.09929",
    archivePrefix = "arXiv",
    primaryClass = "astro-ph.CO",
    month = "8",
    year = "2026"
}

@article{Mingarelli_2017,
   title={The local nanohertz gravitational-wave landscape from supermassive black hole binaries},
   volume={1},
   ISSN={2397-3366},
   url={http://dx.doi.org/10.1038/s41550-017-0299-6},
   DOI={10.1038/s41550-017-0299-6},
   number={12},
   journal={Nature Astronomy},
   publisher={Springer Science and Business Media LLC},
   author={Mingarelli, Chiara M. F. and Lazio, T. Joseph W. and Sesana, Alberto and Greene, Jenny E. and Ellis, Justin A. and Ma, Chung-Pei and Croft, Steve and Burke-Spolaor, Sarah and Taylor, Stephen R.},
   year={2017},
   month=Nov, pages={886–892} }

@article{Gardiner:2023zzr,
    author = "Gardiner, Emiko C. and Kelley, Luke Zoltan and Lemke, Anna-Malin and Mitridate, Andrea",
    title = "{Beyond the Background: Gravitational-wave Anisotropy and Continuous Waves from Supermassive Black Hole Binaries}",
    eprint = "2309.07227",
    archivePrefix = "arXiv",
    primaryClass = "astro-ph.HE",
    doi = "10.3847/1538-4357/ad2be8",
    journal = "Astrophys. J.",
    volume = "965",
    number = "2",
    pages = "164",
    year = "2024"
}

@article{Sato-Polito:2023spo,
    author = "Sato-Polito, Gabriela and Kamionkowski, Marc",
    title = "{Exploring the spectrum of stochastic gravitational-wave anisotropies with pulsar timing arrays}",
    eprint = "2305.05690",
    archivePrefix = "arXiv",
    primaryClass = "astro-ph.CO",
    doi = "10.1103/PhysRevD.109.123544",
    journal = "Phys. Rev. D",
    volume = "109",
    number = "12",
    pages = "123544",
    year = "2024"
}

\end{document}